\documentclass[twoside,leqno,twocolumn]{article}
\usepackage[letterpaper]{geometry}

\usepackage{siamproceedings}

\usepackage[T1]{fontenc}
\usepackage{amsfonts}
\usepackage{graphicx}
\graphicspath{{figures/}}
\usepackage{epstopdf}
\usepackage{enumitem}
\usepackage{algorithmic}
\ifpdf
  \DeclareGraphicsExtensions{.eps,.pdf,.png,.jpg}
\else
  \DeclareGraphicsExtensions{.eps}
\fi
\pdftrue

\usepackage{booktabs}

\newsiamremark{remark}{Remark}
\newsiamremark{hypothesis}{Hypothesis}
\crefname{hypothesis}{Hypothesis}{Hypotheses}
\newsiamthm{claim}{Claim}

\usepackage{amsopn}

\usepackage{xcolor}
\usepackage{enumitem}

\newcommand{\GMR}{\textsc{gmr}}
\newcommand{\IOMR}{\textsc{iomr}}
\newcommand{\DOMR}{\textsc{domr}}
\newcommand{\code}[1]{\texttt{\small #1}}

\newcommand{\restore}{\textsf{restore}}

\newcommand{\loneInfWin}{4.8}
\newcommand{\loneInfN}{419}

\newcommand{\loneDefWin}{100.0}
\newcommand{\loneDefN}{400}

\newcommand{\nmrNAtom}{430}

\newcommand{\nmrCoreAtom}{340}
\newcommand{\nmrDispObsAtom}{0.1489}
\newcommand{\nmrDispAllAtom}{0.1233}

\newcommand{\nmrRelIomrBestofkAtom}{-0.1}

\newcommand{\nmrSMPivotAtom}{75.8}

\newcommand{\nmrRelPivotAtom}{+45.4}

\newcommand{\nmrBestCodeAtom}{\code{iomr\_bestofk}}

\newcommand{\nmrWorstCodeAtom}{\code{pivot}}

\newcommand{\nmrNWinAtom}{5}
\newcommand{\nmrNAlgoAtom}{16}

\newcommand{\nmrCoreResidue}{75}
\newcommand{\nmrDispObsResidue}{0.1047}
\newcommand{\nmrDispAllResidue}{0.0443}

\newcommand{\nmrRelDomrResidue}{+0.0}

\newcommand{\nmrSMLeftEdgeResidue}{51.6}
\newcommand{\nmrRestLeftEdgeResidue}{0.3270}

\newcommand{\nmrBestCodeResidue}{\code{domr}}

\newcommand{\nmrWorstCodeResidue}{\code{left\_edge}}

\newcommand{\nmrNWinResidue}{0}
\newcommand{\nmrNAlgoResidue}{16}

\title{Metric repair is two problems:\\ Which edges, and what weights\thanks{The artifact to the paper can be found here:\\
\href{https://anonymous.4open.science/r/Metric-Repair-Is-Two-Problems-ALENEX/README.md}{Artifact}\\ It will be remain publicly available}.}
\author{Asaf Etgar \and Anna Gilbert}

\begin{document}
\maketitle
\fancyfoot[R]{\scriptsize{Copyright \textcopyright\ 20XX by SIAM\\
Unauthorized reproduction of this article is prohibited}}
\begin{abstract}
Real distance data rarely cooperate: measurements are noisy, observations are missing, and the numbers that result seldom satisfy the triangle inequality. A family of methods exists to correct them, and every one of those methods rests on the same hope --- that analysis run on the corrected data is a more faithful surrogate for the truth than analysis run on the raw data. \emph{We are not aware of anyone having tested that hope.} We test it through the problem of \emph{Metric Repair}, which asks for the fewest edges whose reweighting restores the triangle inequality. We implement a suite of algorithms, covering the literature and new methods, both with theoretical guarantees and heuristics, and evaluate their performance on real and synthetic data, both inherently non metric and corrupted. We demonstrate that the algorithms' performance is determined predominantly by the \emph{type and fraction} of corruption, rather than the corruption's magnitude or graph size. 

We further test the effect of repair on \emph{downstream tasks}, namely MDS and $k$NN, and ask if the repair got the result closer to the truth compared to a corrupted instance. In most cases it did not, and we identify the culprit. A small set of edges is not enough. Finding the correct set of edges, be it an injected corruption or a natural non-metricity, is critical. Moreover, deciding on a weight rule impacts performance: on data instances with available metric ground truth, a metric repair algorithm can pull the graph further from the truth, while an oracle access to the true weights helps. Surprisingly, the opposite can be true as well. Setting the weights is not an implementation detail; it is half the problem.

\end{abstract}

\clearpage
\section{Introduction}\label{sec:intro}
Real distance data is rarely recorded as such. A ping between two hosts is routed, not flown, a similarity turned into a distance by $s \mapsto 1/s$ carries no promise of a triangle inequality, a nuclear magnetic resonance experiment reports an \emph{upper bound} on the separation of two protons, and three such bounds need not be consistent with any arrangement of three points.
The result, in each case, is a weighted graph that a downstream tool will treat as a metric --- and it need not be.

The consumers of such data are unforgiving about it: classical multidimensional scaling~\cite{gilbertMDS,borg2005modern} wants a Euclidean metric and returns spurious geometry when it does not get one, and a nearest-neighbor graph built on distances with fabricated shortcuts contains neighborhoods that do not exist. This is more than a cosmetic defect, and every field that meets this problem has invented a correction for it. Structural biology smooths its distance bounds until they satisfy the triangle inequality before it embeds them \cite{crippenhavel1988,HAVEL1983665}. Network measurement detects the violations, names them, and removes the worst offenders \cite{networkcoordinates}. Complex-networks research strips out the redundant edges entirely and keeps
what it calls the metric backbone \cite{metricbackbone,distancebackbone}, and they share the same underlying assumption:
\emph{ a corrected dataset is a better surrogate for the underlying truth than the raw one.} It is a reasonable assumption, yet as far as we can determine, an untested one: the corrections are validated against self-consistency, not against an \emph{external} ground truth, because an external ground truth is usually unavailable.

We assemble a collection on which it \emph{is} available --- synthetic metric graphs with injected corruptions, a protein whose structure is known, a set of internet hosts whose coordinates are known, a road network whose geography is known --- and we measure whether the correction helps.

The correction we study is \emph{metric repair}: find a smallest set $S \subseteq E$ whose reweighting makes
the graph metric. An active combinatorial literature has studied it \cite{gilbert2017sparse,fan_et_al:LIPIcs.SWAT.2020.25,metricsultrametrics,fan2018metric}, separating three variants (\DOMR{}, \IOMR{} and \GMR{}). We note that  \DOMR{} unifies the three corrections above. But an algorithm for metric repair returns a \emph{set} $S$, while a tool consumes a \emph{graph} --- a distance matrix, an embedding,
a neighborhood --- and to obtain one, the edges of $S$ must be given \emph{weights}. The literature supplies a single rule:
reweigh each edge in $S$ to the shortest path that avoids $S$. We call it \restore{}.

The rule is so natural that the literature does not present it as a choice at all. It is folded into the statement of the problem, and nobody asks whether it is right. \emph{We ask; It is not.}

There are several significant fields of applied and computational mathematics in which the essential difficulty lies in identifying the support of the unknown signal: once the correct set is known, the coefficient values are recovered easily. Accordingly, a substantial line of work in compressive sensing treats exact support recovery as the central success criterion and characterizes precisely when it is achievable~\cite{Wainwright2009Sharp}, while greedy algorithms such as orthogonal matching pursuit are designed and analyzed explicitly as support-identification procedures~\cite{TroppGilbert2007OMP}. An analogous phenomenon appears in classical inverse problems: in inverse scattering, qualitative methods aim only to determine the support of the unknown scatterer, after which recovering material parameters on the identified region is comparatively benign~\cite{Kirsch1998Factorization,Cheney2001LSM}. Our problem is fundamentally different.

\noindent\textbf{Contributions.} 
\begin{itemize}[leftmargin=*,itemsep=1pt,topsep=2pt]
\item We implement a suite of algorithms and heuristics for metric repair both from the literature and novel: exact solvers, LP relaxations and rounding, combinatorial approximation algorithms. 
\item We test these algorithms on a broad set of example graphs, both synthetic and real data, metric and non-metric. We explore two directions of corruption on metric graphs, and change their magnitude and fraction of corruption. We identify that the deciding factor in algorithms' performance is the corruption's \emph{direction and fraction}, rather than its magnitude or graph size. No method dominates all regimes, and so no method can be recommended ex ante.
\item The real world datasets we test cover the three communities already exercising metric repair algorithms, and include both metric and non-metric examples. We test the impact of metric repair on tasks that are routinely performed on graphs that are assumed to be metric, namely $k$NN and MDS. We find that the effect of metric repair on downstream tasks is mostly \emph{negative}.  On five real graphs with an external ground truth, no repair in the suite improves $k$-NN recovery at any $k$. Geometry improves on one graph of five. We say where, and why, and where it does not. They correct the data to a metric, but not a metric that resembles the truth. While they can identify an injected corruption, the repair they offer is either on the wrong edges or using the wrong weights and the two are intertwined. 
\end{itemize}
Consequently, we find that metric repair requires more delicate treatment than finding the smallest set of edges. Finding a small support is not the only deciding factor in the quality of repair. Give a good, small set of edges the true weights and it recovers $k$NN and MDS. Give the same set the weights used in the literature and it recovers nothing.
The right set of edges is not much without the right weight rule, and a good rule is not enough without knowing which edges hold the corruption. A weight rule should be \emph{certified}, make the graph metric; \emph{recovering}, move toward the truth; and \emph{oblivious}, use only the observed graph. The literature uses a rule that is certified and oblivious; having oracle access to the true weights is certified and recovering; we know of nothing with all three.

\subsection{Related work: one problem, three communities}\label{sec:related}

Metric repair on complete graphs was introduced independently by Gilbert and Jain ~\cite{gilbert2017sparse} and by Fan, Raichel and Van~Buskirk~\cite{fan2018metric}.  Fan, Raichel and Van~Buskirk showed that it is $\mathrm{NP}$---hard and provided the first approximation algorithm; Cohen-Addad, Fan, Lee and de Mesmay~\cite{metricsultrametrics} provided an $O(\log(n))$ approximation algorithm and a better algorithm for \emph{ultrametrics}; Fan, Gilbert, Raichel, Sonthalia and Van~Buskirk~\cite{fan_et_al:LIPIcs.SWAT.2020.25} generalized the problem to non-complete graphs and provided approximation algorithms as well as a first FPT algorithm. Every one of these poses the \emph{set} problem --- find the fewest edges ---
and if setting weights is discussed, use \restore{}.

The same problem is solved, under other names, by fields that do not cite one another. Structural biology performs \emph{bound smoothing} on distance-geometry data, and Havel's smoothed upper limits \emph{are} the all-pairs shortest paths of the bound graph~\cite{HAVEL1983665} --- exactly our \code{observed} baseline. Complex-networks research strips \emph{semi-metric} edges to keep the \emph{metric backbone}~\cite{metricbackbone,distancebackbone} --- our $H$ and our \DOMR{}. Network measurement detects \emph{triangle-inequality violations} and discards the
worst~\cite{networkcoordinates} --- our heavy set again. Three communities, three special cases, one problem; and none of
them asks what the repaired weights should be. Adjacent and already settled is the $\ell_0$ fit to a \emph{tree}
metric~\cite{treemetric}, a strictly stronger requirement than the triangle inequality.

\subsection{AI declaration}
We used Claude Opus 4.8 to help translate code written in Sagemath to Python and write SLURM submission scripts for experiments.
\section{Preliminaries}\label{sec:prelim}
%
%

A graph $G = (V, E, w)$ is a connected undirected graph with positive edge
weights $w: E \to \mathbb{R}_{>0}$, on $n = |V|$ vertices and $m = |E|$ edges. We write $d_G(u,v)$ for
the length of a shortest $u$--$v$ path, and $G \setminus S$ for the graph with the edges of $S \subseteq E$
deleted. A shorthand we use throughout: for an edge $e = uv$, its \emph{detour}
\[
  \delta_G(e) = d_{G \setminus e}(u,v)
\]
is the shortest way between its endpoints that does not use it.

An edge $e = uv$ is \emph{heavy} if $w(e) > \delta_G(e)$, and we write
\[
  H(G) \;=\; \{\, e \in E \;:\; w(e) > \delta_G(e) \,\}
\]
for the \emph{heavy set}. The graph is \emph{metric} exactly when $H(G) = \emptyset$; that is, when every
edge is a shortest path between its own endpoints. Call a cycle in $G$ \emph{broken} if it contains an edge $e\in H\cap C$ and $w(e) > d_G(C\setminus e)$.
 
\noindent
\textbf{Metric repair.} Given a non-metric $G$, a \emph{repair} is a pair $(S, w')$: a set
$S \subseteq E$ of edges and an assignment $w': S \to \mathbb{R}_{>0}$ of new weights to them, such that setting the weights of $S$ to $w'$ in $G$ turns it metric. The optimization problem the literature poses is to minimize $|S|$. 

The variants in the literature differ in what
$w'$ is permitted to do to every edge $e \in S$:
\begin{itemize}[leftmargin=*,itemsep=1pt,topsep=2pt]
\item \DOMR{} (decrease-only): $w'(e) \le w(e)$ .
\item \IOMR{} (increase-only): $w'(e) \ge w(e)$.
\item \GMR{} (general): $w'$ is unconstrained.
\end{itemize}
Every one of them is a constraint on the \emph{weights}, and every one of them is nevertheless scored on the
\emph{set}. Given $S$, a suitable $w'$ exists if and only if $S$ intersects every broken cycle in $G$~\cite[Theorem 2]{fan_et_al:LIPIcs.SWAT.2020.25}, with some restrictions on $S$ for the \DOMR{} and \IOMR{} variants. Therefore, we call $S$ a \emph{hitting set}. We mention that \restore{} is the $w'$ used in the proof of this theorem.

The \DOMR{} problem is considered closed in the literature:

\begin{lemma}[\cite{fan_et_al:LIPIcs.SWAT.2020.25}]\label{lem:backbone}
$H(G)$ is an optimal \DOMR{} cover, and is computable using all pairs shortest path.
\end{lemma}
As a consequence, we get the following result important to our discussion:
\begin{lemma}[decrease-only invariance; \cite{gilbert2017sparse,fan_et_al:LIPIcs.SWAT.2020.25}]\label{lem:noop}\label{lem:domr}
If $w(e) \ge d_{G \setminus S}(e)$ for every $e \in S$, then applying \restore{} to $S$ leaves every
shortest-path distance unchanged.
\end{lemma}

\restore{} is a choice. It is the \emph{largest} value that
keeps $e$ from being heavy, so it is the smallest edit that certifies metricity. We give it a name so that we can ask whether it is the right one.







\subsection{The corruption model}\label{sec:corrupt}

Our synthetic instances are built by breaking a graph (in two possible ways) that is metric by construction, so the corrupted set is
known. Both corruptions take a metric $G$, a \code{fraction} $q \in (0,1)$ and a \code{magnitude} $\mu > 1$, and draw a set
$Q \subseteq E$ of $\lceil qm \rceil$ edges uniformly at random. They differ in what they do to an edge, and the difference is the axis on which most of this paper turns.

\begin{itemize}[leftmargin=*,itemsep=1pt,topsep=2pt]
\item \emph{Inflate:} For $e = uv \in Q$ lying on a cycle,
\[
  w'(uv) \;\gets\; \mu \cdot \delta_G(uv).
\]

\item \emph{Deflate: }For $e = uv \in Q$, with at least one common neighbor, 
set
\(
  \mathrm{gap}(uv)= \max_{c \in N} \bigl| \, w(uc) - w(cv) \, \bigr|,
\)
where $N$ is the common neighborhood of $u,v$,
and set
\[
  w'(uv) \;\gets\; \mathrm{gap}(uv) \, / \, \mu .
\]
\end{itemize}





The two corruptions are not mirror images. Under inflation the heavy set is (a subset of) the corrupted set, $H \subseteq B$, so by
Lemma~\ref{lem:backbone} the decrease-only optimum finds the culprits in polynomial time. Under deflation $H$ holds the \emph{victims} of the corruption: a deflated edge turns one of its neighbors heavy. Whatever an algorithm finds when it hunts for heavy edges, under deflation it is not hunting for the damage.

Note that $q$ is a \emph{request}, not a
guarantee: an edge on no cycle cannot be inflated, and an edge with no common neighbor has no $2$-path to
undercut, so both corruptions skip what they cannot break. We report the realized $|B|$, not
$\lceil q\,m \rceil$. In the synthetic families we break, $|B| \approx q\cdot m$ approximately $99\%$ of the time. We report the percentage on the real dataset we break. Second, inflation measures each detour in the \emph{clean} graph but heaviness in the
\emph{corrupted} one, so when two inflated edges share a detour a weak inflation need not survive its own
interference: $H \subseteq B$, with equality occurring more frequently as $\mu$ grows and $q$ shrinks. Thus \DOMR{} recovers the corrupted set with precision $1$ but recall $|H|/|B| \le 1$.



\section{Setup}

\subsection{Algorithms}
We benchmark a suite spanning the three variants and the full range of guarantees. The solvers are divided into four groups:

\begin{itemize}[leftmargin=*,itemsep=1pt,topsep=2pt]
\item Exact Solvers: \code{domr} solves the \DOMR{} problem according to \ref{lem:domr}. \code{gmr\_ilp,iomr\_ilp} are integer programs that solve \GMR{} and \IOMR{} exactly if they converge. 
\item Combinatorial Solvers: \code{pivot}~\cite{treemetric},\code{left\_edge}~\cite{gilbert2017sparse} are the algorithms for metric repair on the complete graph. We complete the graph, run these algorithms, and only keep the edges from the original graph that they return. \code{spc\_gmr} and \code{spc\_iomr} are the Shortest Path Cover algorithms from~\cite{fan_et_al:LIPIcs.SWAT.2020.25}.
\item $\ell_1$ minimizers: \code{l1sep\_gmr} and \code{l1sep\_iomr} solve a linear program to minimize the total sum of weights changed to turn the graph metric. We keep only the edges changed, not the changed weights.
\item LP solvers: the rest of the algorithms solve the hitting-set LP relaxation for metric repair, and round the solution in different ways (randomized, threshold, region growing). 
\end{itemize}
All LPs and ILPs use cutting-plane (separation oracle) implementations. The full details are provided in the appendix~\ref{tab:suite}
\subsection{Synthetic families}\label{sec:families}
\begin{table}[t]\centering\footnotesize\setlength{\tabcolsep}{3pt}
\caption{The real datasets. $|H|/m$ is the heavy-edge fraction; $6$ of the $10$ graphs carry an external ground truth, and only those can say whether a repair moves a graph \emph{toward} anything. The last column asks whether a \emph{standard} downstream tool in that field needs the triangle inequality at all: for $8$ of the $10$ the answer is no. On the similarity graphs the non-metricity is manufactured by the distance conversion: on \code{flycns\_male}, $|H|/m$ swings from $0.6\%$ to $83.1\%$ with the choice alone.}
\label{tab:datasets}
\begin{tabular}{@{}lrrrcc@{}}
\toprule
graph & $n$ & $m$ & $|H|/m$ (\%) & truth & downstream \\
\midrule
\code{nmr\_atom} & $430$ & $1{,}357$ & $1.1$ & $\checkmark$ & \textbf{yes} \\
\code{nmr\_res} & $75$ & $308$ & $5.2$ & $\checkmark$ & \textbf{yes} \\
\code{dimacs\_ny\_d} & $5{,}000$ & $6{,}017$ & $0.0$ & $\checkmark$ & n/a \\
\code{dimacs\_ny\_t} & $5{,}000$ & $6{,}017$ & $0.3$ & $\checkmark$ & n/a \\
\code{ripe\_atlas} & $999$ & $442{,}707$ & $95.3$ & $\checkmark$ & no \\
\code{pbmc3k} & $2{,}700$ & $31{,}639$ & $0.2$ & $\checkmark$ & no \\
\code{cassiopeia} & $1{,}000$ & $12{,}760$ & $30.7$ & \code{--} & no \\
\midrule
\code{bct\_coact} & $638$ & $18{,}625$ & $1.1$--$43.5$ & \code{--} & no \\
\code{flycns} & $1{,}200$ & $14{,}025$ & $0.6$--$83.1$ & \code{--} & no \\
\code{fish1\_ten} & $1{,}000$ & $1{,}175$ & $0.0$--$2.0$ & \code{--} & no \\
\bottomrule
\end{tabular}
\end{table}

We design two synthetic families of random graphs, one that is inherently metric and one that is inherently not. 

\subsubsection{Random Geometric Graphs.} The metric family is a Random Geometric Graph~\cite{penrose2003random}: draw $n$ points uniformly in the unit square and join two whenever they lie within a radius $r$, weighting each edge by the Euclidean distance between its endpoints. We set $r = \sqrt{\mathrm{deg}/(\pi n)}$ to fix the expected degree, and use $\mathrm{deg} = 12$ unless we say otherwise.

Two properties make RGGs suitable. They are \emph{sparse}, so an algorithm that ignores sparsity has somewhere to fail, and their density and weight model are \emph{independent}: the radius sets one and the geometry sets the other. 
\paragraph{Corruption knobs}
A metric synthetic family gives us something real data often cannot: a corrupted set we \emph{know}. We turn three knobs:
\begin{itemize}[leftmargin=*,itemsep=1pt,topsep=2pt]
\item the \textbf{direction} --- inflate or deflate.
\item the \textbf{fraction} $q$, the share of edges drawn for corruption (baseline $0.10$);
\item the \textbf{magnitude} $\mu$, how far past the break threshold each edge is pushed (baseline $3.0$).
\end{itemize}

Remember that $q$ is a request, not a guarantee (see \S\ref{sec:corrupt}). On the RGG the request is satisfied: $99.7\%$ of the time, because a geometric graph is heavily clustered and $99.5\%$ of its edges sit on a triangle. On a road network
it does not --- $28.7\%$ of the edges are bridges, and a requested $10\%$ deflation reaches $6\%$.


\subsection{Real datasets}
We collect nineteen real distance graphs from molecular structure (NMR~\cite{nmr1d3z}), internet latency
(RIPE Atlas~\cite{ripeatlas}), road networks (DIMACS~\cite{dimacs9}), phylogenetics~\cite{cassiopeia},
connectomics~\cite{coactivation} and single-cell genomics~\cite{pbmc3k}; Table~\ref{tab:datasets} gives
their sizes and non-metric fractions. They differ in an important way we exploit: whether a standard downstream tool actually requires a metric, and whether an external ground truth exists against which repair can be judged. Only a few
meet both criteria, and we are explicit about which.

\subsection{Experimental Details}
Experiments run on shared-cluster nodes with 128-core AMD EPYC 9575F processors with 2.2TB RAM per node. Algorithms were implemented in python with standard packages (numpy, scikit). See \href{https://anonymous.4open.science/r/Metric-Repair-Is-Two-Problems-ALENEX/README.md}{\underline{The Artifact}} for full details.
\section{Algorithmic Benchmarks}\label{sec:benchmarks}

%
%
\newcommand{\secLandSmallDeflate}{99.7}
\newcommand{\secLandWorstSmallDeflate}{93.8}
\newcommand{\secLandSmallInflate}{99.8}
\newcommand{\secLandWorstSmallInflate}{83.3}
\newcommand{\secLandLargeDeflate}{99.7}
\newcommand{\secLandWorstLargeDeflate}{98.6}
\newcommand{\secLandLargeInflate}{99.6}
\newcommand{\secLandWorstLargeInflate}{88.7}
\newcommand{\secSmallTasks}{2{,}960}
\newcommand{\secSmallNLo}{100}
\newcommand{\secSmallNHi}{500}
\newcommand{\secSmallJitterDropped}{0}
\newcommand{\secGmrIlpSolved}{2{,}837}
\newcommand{\secGmrIlpPct}{96}
\newcommand{\secIomrIlpSolved}{1{,}476}
\newcommand{\secIomrIlpPct}{50}
\newcommand{\secInflateTasks}{200}
\newcommand{\secInflateOptOverH}{1.000}
\newcommand{\secInflateHOverB}{1.00}
\newcommand{\secDeflateTasks}{120}
\newcommand{\secDeflateOptOverH}{0.209}
\newcommand{\secDeflateHOverB}{4.73}
\newcommand{\secInflateRecall}{99.8}
\newcommand{\secInflateExact}{76.5}
\newcommand{\secOptLoneGmrInflate}{5.006}
\newcommand{\secOptLoneGmrDeflate}{1.262}
\newcommand{\secOptSpcGmrInflate}{2.582}
\newcommand{\secOptSpcGmrDeflate}{6.859}
\newcommand{\secOptPivotInflate}{7.110}
\newcommand{\secOptPivotDeflate}{4.945}
\newcommand{\secOptLeftEdgeInflate}{2.532}
\newcommand{\secOptLeftEdgeDeflate}{4.545}
\newcommand{\secOptLoneIomrInflate}{1.489}
\newcommand{\secOptLoneIomrDeflate}{1.263}
\newcommand{\secMeanPivotInflate}{7.09}
\newcommand{\secSdPivotInflate}{0.20}
\newcommand{\secMeanPivotDeflate}{4.94}
\newcommand{\secSdPivotDeflate}{0.19}
\newcommand{\secMeanLeftEdgeInflate}{2.52}
\newcommand{\secSdLeftEdgeInflate}{0.10}
\newcommand{\secMeanLeftEdgeDeflate}{4.54}
\newcommand{\secSdLeftEdgeDeflate}{0.18}
\newcommand{\secMeanLoneGmrInflate}{4.99}
\newcommand{\secSdLoneGmrInflate}{0.20}
\newcommand{\secMeanLoneGmrDeflate}{1.26}
\newcommand{\secSdLoneGmrDeflate}{0.04}
\newcommand{\secNErratic}{0}
\newcommand{\secLadderNLo}{1000}
\newcommand{\secLadderNHi}{3000}
\newcommand{\secSmLoneGmrInflate}{0.508}
\newcommand{\secSmLoneGmrDeflate}{0.124}
\newcommand{\secSmSpcGmrInflate}{0.257}
\newcommand{\secSmSpcGmrDeflate}{0.683}
\newcommand{\secInflateAboveH}{12}
\newcommand{\secInflateNAlgos}{12}
\newcommand{\secDeflateAboveH}{3}
\newcommand{\secDeflateNAlgos}{12}
\newcommand{\secTopN}{3000}
\newcommand{\secTopM}{17{,}444}
\newcommand{\secTopCompletion}{4{,}498{,}500}
\newcommand{\secTopBlowup}{258}
\newcommand{\secMemPivot}{2{,}347}
\newcommand{\secMemLeftEdge}{2{,}306}
\newcommand{\secMemDomr}{878}
\newcommand{\secMemRestLo}{1.01}
\newcommand{\secMemRestHi}{1.05}
\newcommand{\secMemRestPct}{5}
\newcommand{\secMemPivotRel}{2.6}
\newcommand{\secMemLeftEdgeRel}{2.5}
\newcommand{\secMemBoundRel}{0.53}
\newcommand{\secMemDomrLo}{165}
\newcommand{\secMemDomrHi}{878}
\newcommand{\secTopWorstTimeoutAlgo}{\code{gmr\_bestofk}}
\newcommand{\secTopWorstTimeoutPct}{50}
\newcommand{\secRgrowHmax}{200}
\newcommand{\secRgrowTopRet}{0}
\newcommand{\secRetLoneGmrInflate}{95}
\newcommand{\secRetSpcGmrInflate}{100}
\newcommand{\secRetGmrBestofkInflate}{0}
\newcommand{\secRetGmrRandInflate}{100}
\newcommand{\secRetGmrThrInflate}{100}
\newcommand{\secRetPivotInflate}{100}
\newcommand{\secRetLoneIomrInflate}{90}
\newcommand{\secRetSpcIomrInflate}{100}
\newcommand{\secRetIomrBestofkInflate}{0}
\newcommand{\secRetIomrRandInflate}{100}
\newcommand{\secRetIomrThrInflate}{100}
\newcommand{\secRetIomrRgrowInflate}{0}
\newcommand{\secRetLeftEdgeInflate}{100}
\newcommand{\secNDeadInflate}{2}
\newcommand{\secRetLoneGmrDeflate}{100}
\newcommand{\secRetSpcGmrDeflate}{100}
\newcommand{\secRetGmrBestofkDeflate}{100}
\newcommand{\secRetGmrRandDeflate}{100}
\newcommand{\secRetGmrThrDeflate}{100}
\newcommand{\secRetPivotDeflate}{100}
\newcommand{\secRetLoneIomrDeflate}{100}
\newcommand{\secRetSpcIomrDeflate}{100}
\newcommand{\secRetIomrBestofkDeflate}{100}
\newcommand{\secRetIomrRandDeflate}{100}
\newcommand{\secRetIomrThrDeflate}{100}
\newcommand{\secRetIomrRgrowDeflate}{0}
\newcommand{\secRetLeftEdgeDeflate}{100}
\newcommand{\secNDeadDeflate}{0}
\newcommand{\secCycInflate}{15.7}
\newcommand{\secHsetInflate}{1{,}741}
\newcommand{\secCycDeflate}{1.2}
\newcommand{\secHsetDeflate}{7{,}442}
\newcommand{\secCycRatio}{13}
\newcommand{\secPooledWin}{51.3}
\newcommand{\secPooledShare}{48.8}
\newcommand{\secPooledInfWin}{4.8}
\newcommand{\secPooledDefWin}{100}
\newcommand{\secPooledN}{819}
\newcommand{\secMatchN}{300}
\newcommand{\secTimeoutCap}{1800}
\newcommand{\secFMDirs}{deflate}
\newcommand{\secFMOneDirection}{true}
\newcommand{\secFMSweepN}{1000}
\newcommand{\secFracLo}{0.02}
\newcommand{\secFracHi}{0.3}
\newcommand{\secFracLoneGmrLo}{0.023}
\newcommand{\secFracLoneGmrHi}{0.362}
\newcommand{\secFracSpcGmrLo}{0.259}
\newcommand{\secFracSpcGmrHi}{0.952}
\newcommand{\secFracHmLo}{0.148}
\newcommand{\secFracHmHi}{0.633}
\newcommand{\secMagLo}{2}
\newcommand{\secMagHi}{10}
\newcommand{\secMagLoneGmrLo}{0.120}
\newcommand{\secMagLoneGmrHi}{0.125}
\newcommand{\secMagSpcGmrLo}{0.579}
\newcommand{\secMagSpcGmrHi}{0.824}
\newcommand{\secMagHmLo}{0.333}
\newcommand{\secMagHmHi}{0.582}

We benchmark the suite on the planted RGG family of Section~\ref{sec:families},  a metric graph whose true weights $w_0$ we keep and whose corrupted set $B$ we know. 
Over this wide range of instances, we can see that no one algorithm dominates. Which one is the best depends on both the type of instance and the type of corruption. While the ranking of an algorithm depends on the input graph and corruption types, the relative position does not depend on whether the figure of merit is the output size compared to the optimal size or simply the total number of edges, thus giving us a comparison metric even in the large graph cases or those for which we have no optimal solution. We find that the memory is dominated by APSP after finding a hitting set and that running time does not seem to be a factor in ranking algorithms. 
\subsection{Known $\mathrm{OPT}$.}\label{sec:s51}
On small instances, the exact ILP solvers converge so we can measure each algorithm against the true optimum. On the sparse family ($n = \secSmallNLo{}$--$\secSmallNHi{}$,
\secSmallTasks{} planted instances) a cutting-plane
integer program solves $\secGmrIlpPct{}\%$ of the \GMR{} instances;
for \IOMR{} the ILP converges on, $\secIomrIlpPct{}\%$ of the instances.

When we consider RGGs across the two types of corruption, \emph{no algorithm is a clear winner}.  \code{spc\_gmr} is the best \GMR{} heuristic under inflation
($\secOptSpcGmrInflate{}\times$ optimal) and the worst under deflation ($\secOptSpcGmrDeflate{}\times$);
\code{l1sep\_gmr} does the reverse ($\secOptLoneGmrInflate{}\times$ / $\secOptLoneGmrDeflate{}\times$).

Under inflation, the heavy set is exactly the \GMR{} optimum: $\mathrm{OPT}/|H| = \secInflateOptOverH{}$, an inflated edge exceeds its own detour, so every heavy edge must be adjusted, and editing exactly the heavy edges is a feasible hitting set with none smaller. The heavy set is \emph{almost} the corrupted set, but not necessarily always: $H \subseteq B$ always (precision $1$), with recall $\secInflateRecall{}\%$ and equality $H = B$ on $\secInflateExact{}\%$ of instances. The discrepancy between the two is the weakly inflated edges; those edges whose detour runs through another inflated edge such that the detour grows too and the planted edge is no longer in violation. Under deflation, the heavy set is neither optimum nor necessarily the candidate set; it contains the edges impacted by the shortcuts, and runs $\secDeflateHOverB{}\times$ the size of the corrupted set. No heuristic beats it under inflation; none needs it under deflation.

On large synthetic instances (beyond $n \approx \secSmallNHi{}$) the integer programs no longer converge and the optimum is gone; we report $|S|/m$, the \emph{fraction of the graph} a repair rewrites. $|S|/\mathrm{OPT}$ and $|S|/m$ differ only by the per-instance factor $\mathrm{OPT}/m$, so they induce the same ordering within each instance; we report $|S|/m$ because it is defined at every scale. Wherever the graph has something to repair, the two rank the algorithms the same way (Table~\ref{tab:optm}). 

\subsection{Where the algorithms break.}\label{sec:s52}
On larger instances, the inversion phenomena remain. \code{l1sep\_gmr} repairs
$\secSmLoneGmrDeflate{}$ of the graph under deflation and $\secSmLoneGmrInflate{}$ under inflation --- four-fifths of every edge --- while \code{spc\_gmr} runs the other way ($\secSmSpcGmrInflate{}$ /
$\secSmSpcGmrDeflate{}$). We report the ranking in Table~\ref{tab:invert}. The direction decides not just the ranking but which methods run at all: under inflation \secNDeadInflate{} of them exceed the $\secTimeoutCap{}$\,s cap and return nothing (the covering-LP
roundings), against \secNDeadDeflate{} under deflation. A timeout is a result, not simply a missing value. 
\begin{table}[ht]\centering\footnotesize
\caption{\textbf{The median share of the graph a repair rewrites}. Each column is a corruption type and we only consider runs that returned \emph{and} verified turning the graph metric; \emph{ret.} is how often a verified hitting set came back at all (420 + 400 tasks). \DOMR{}'s row \emph{is} $|H|/m$. Bold: best and two worst \emph{within} each variant and direction. A dash is no hitting set. }

\label{tab:invert}
\resizebox{\columnwidth}{!}{%
\begin{tabular}{@{}lrr@{\quad}rr@{}}
\toprule
 & \multicolumn{2}{c}{inflate} & \multicolumn{2}{c}{deflate} \\
\cmidrule(lr){2-3} \cmidrule(lr){4-5}
algorithm & ret. & $|S|/m$ & ret. & $|S|/m$ \\
\midrule
\addlinespace[1pt]
\multicolumn{5}{@{}l}{\DOMR{}} \\[1pt]
\quad \code{domr} & $100.0\%$ & $0.100$ & $100.0\%$ & $0.429$ \\
\midrule
\multicolumn{5}{@{}l}{\GMR{}} \\[1pt]
\quad \code{spc\_gmr} & $100.0\%$ & $\mathbf{0.256}$ & $100.0\%$ & $\mathbf{0.685}$ \\
\quad \code{gmr\_bestofk} & $43.8\%$ & $0.324$ & $100.0\%$ & $0.136$ \\
\quad \code{gmr\_rand} & $100.0\%$ & $0.345$ & $100.0\%$ & $0.137$ \\
\quad \code{gmr\_thr} & $100.0\%$ & $0.345$ & $100.0\%$ & $0.137$ \\
\quad \code{l1sep\_gmr} & $99.8\%$ & $\mathbf{0.510}$ & $100.0\%$ & $\mathbf{0.124}$ \\
\quad \code{pivot} & $100.0\%$ & $\mathbf{0.807}$ & $100.0\%$ & $\mathbf{0.526}$ \\
\midrule
\multicolumn{5}{@{}l}{\IOMR{}} \\[1pt]
\quad \code{l1sep\_iomr} & $99.5\%$ & $\mathbf{0.510}$ & $100.0\%$ & $\mathbf{0.124}$ \\
\quad \code{iomr\_bestofk} & $55.0\%$ & $0.558$ & $100.0\%$ & $0.125$ \\
\quad \code{iomr\_rand} & $100.0\%$ & $0.566$ & $100.0\%$ & $0.126$ \\
\quad \code{iomr\_thr} & $100.0\%$ & $0.566$ & $100.0\%$ & $0.126$ \\
\quad \code{spc\_iomr} & $100.0\%$ & $\mathbf{0.625}$ & $100.0\%$ & $\mathbf{0.258}$ \\
\quad \code{left\_edge} & $100.0\%$ & $\mathbf{0.842}$ & $100.0\%$ & $\mathbf{0.465}$ \\
\quad \code{iomr\_rgrow} & $0.0\%$ & \code{--} & $0.0\%$ & \code{--} \\
\bottomrule
\end{tabular}
}
\end{table}

The one resource limit is memory. Every method that produces a hitting set first runs one all-pairs shortest path (APSP) pass --- \DOMR{} \emph{is} that pass --- and peaks within $\secMemRestLo{}$--$\secMemRestHi{}\times$
of it. The APSP computation dominates the memory cost. \code{pivot} and \code{left\_edge}, which \emph{complete the graph}: they pay $\Theta(n^2)$
whether the edges are there or not, so at $n = \secTopN{}$ they build $\secTopCompletion{}$ edges for a graph of
$\secTopM{}$. 
It is the same defect that makes them the worst hitting sets in the study. 

When we plant a corruption, we have two parameters under our control, \code{fraction} and \code{magnitude}, and they are not equal. These sweeps run under $\secFMDirs{}$ only at $n = \secFMSweepN{}$: raising the corrupted fraction from $\secFracLo{}$ to $\secFracHi{}$ drives \code{l1sep\_gmr}'s hitting set from $\secFracLoneGmrLo{}$ to $\secFracLoneGmrHi{}$ of the graph, while the magnitude, swept from $\secMagLo{}\times$ to $\secMagHi{}\times$, barely stirs it ($\secMagLoneGmrLo{}$ to $\secMagLoneGmrHi{}$). In other words, fraction is a dial; magnitude is not. Both, importantly, leave the ranking almost untouched. Figure~\ref{fig:fracmag} draws both sweeps.

Table~\ref{tab:invert} deliberately keeps the corruption type reports separate: \code{l1sep\_gmr} returns a smaller hitting set than 
\code{spc\_gmr} on $\loneInfWin{}\%$ of the \loneInfN{} inflation-corrupted instances and $\loneDefWin{}\%$ of the
\loneDefN{} deflation ones, and similar differences exist between other pairs of algorithms. Any average over the two columns would hide this difference in performance.
Choosing the best algorithm therefore has a \emph{conditional} answer: under a certain corruption or certain family, one algorithm prevails. Change one of the two, and another wins. Nothing in an observed graph announces which corruption produced it, so no method is safe to recommend blind. The hitting set, in every case, is only half of a repair --- and we have not yet said a word about
the weights.

\section{Repair evaluation via post-processing}\label{sec:weights}
A repair is a set \emph{and} an assignment of weights (Section~\ref{sec:prelim}), and the literature addresses only the first item. Having chosen a repair set, one must give its edges \emph{numbers}, and every method uses \restore{}. Since we cannot recommend one algorithm blindly, we ask from this point on which algorithm performs best \emph{on the task at hand}. 

We introduce terminology we use throughout. 
\subsection{Testing vehicles}
To check what repair with the naive \restore{} weights assignments buys down the line, we check two tasks. 
\paragraph{$k$NN (topology).} Given a weighted graph $(G=(V,E),w)$ and an integer $k$, for each vertex $v$ write $D_w(v)$ for the set of the $k$ closest vertices to $v$ under the weight function $w$. If $\tilde{w}$ is a different weight function, the \emph{Jaccard similarity} between $w,\tilde{w}$ is 
\[
\frac{1}{|V|}\sum_{v\in V} \frac{|D_w(v)\cap D_{\tilde{w}}(v)|}{|D_w(v)\cup D_{\tilde{w}}(v)|}.
\] We will think of $w$ as the true metric on the graph $G$, and $\tilde{w}$ as either the corrupted weight function or the weight function after a repair algorithm. This is a similarity measure, so higher is better.
\paragraph{MDS (geometry).} Given a weighted graph $G$ and a weight function $w$, it induces a natural distance matrix $D_w$ where the $i,j$ entry contains the square of the $w$-weighted distance between the vertices $i$ and $j$. Classical Multidimensional scaling (cMDS) takes a distance matrix $D_w$ squared entry-wise and finds Euclidean coordinates $P\in \mathbb{R}^{n\times d}$ that best represent the distances $D_w$. We mention here that we run two versions of MDS: cMDS and SMACOF, and the results are consistent with both. We report the cMDS ones. The specifics are out of scope for this paper, and we refer the reader to~\cite{borg2005modern}. $d$ is a dimension of choice for the embedding, and we will use $2$ or $3$ when necessary. 

Given two positions $P,P'$, the \emph{Procrustes disparity} between them is 
\[
||P - QP'||_F
\]
where $||\cdot||_F$ is the Frobenius norm and $Q$ is an orthogonal projection that minimizes this quantity. This is a disparity measure, so lower is better. 

To assess the quality of repair, we compare the repaired graph $(G,w')$ with a graph representing the ground truth: Either the uncorrupted RGG for synthetic experiments, or a reasonable ground truth when considering real data. 

When processing and analyzing graph data, one needs to make sure the vertex labeled are maintained. When comparing the \emph{quality} of two processed graphs, this invariant is even more important: If $G$ is a graph with two weight functions $w,\widetilde{w}$, the \emph{unlabeled} $k$NN structure induced by $w$ and $\widetilde{w}$ could be identical, but the \emph{labeled} structure is completely wrong. 
We make sure the quantities (Jaccard, Procrustes) measure \emph{labeled relationships}, and not the properties of the unlabeled underlying graph. Denote by $\Phi(\cdot,\cdot)$ either Jaccard or Procrustes measure. If $D,\widetilde{D}$ are the two  distance matrices that come from $w,\widetilde{w}$ where rows and columns are consistently indexed by vertices, then $\Phi(D,\widetilde{D})$ is the quantity of interest. If $\pi(D)$ is a permutation of rows and columns of $D$, we verify that $\Phi(\pi(D),\widetilde{D})$ collapses to its worst possible. That is, $\Phi$ does not read information from the unlabeled graph, it makes sure the labels are preserved.

\begin{table}[t]\centering\footnotesize\setlength{\tabcolsep}{2.5pt}\renewcommand{\arraystretch}{1.0}
\caption{\textbf{The corruption determines geometry/topology performance.} Geometry and topology quality change, each as a multiplier of the observed graph's value. \emph{best} is the algorithm with the largest improvement, and \emph{median} is the median improver. On the RGG, deflation wins both axes and inflation loses both --- the median algorithm making the map $8.1\times$ worse while keeping only $0.72$ of the neighborhoods; on the road network the two axes disagree, inflation improving the geometry while degrading the topology. Sizes: planted RGG $n = 3{,}000$, \code{dimacs\_ny\_d} $n = 5{,}000$; $k$NN were computed with $k = 20$. \emph{ret} is how many of the 14 repair algorithms returned within the time cap of $2$ hours.}
\label{tab:corruption}
\resizebox{\columnwidth}{!}{%
\begin{tabular}{@{}llrrrr|rr@{}}
\toprule
graph & corruption & ret & obs. & best & vs.\ obs. & median & vs.\ obs. \\
\midrule
\multicolumn{8}{@{}l}{\textbf{geometry: Procrustes disparity ($\downarrow$; ${<}1$ is better)}} \\[2pt]
planted RGG & deflation & $14$ & $0.0116$ & \shortstack[r]{{\tiny\texttt{spc\_iomr}}\\[-1pt]$0.0024$} & $\mathbf{0.21\times}$ & \shortstack[r]{{\tiny\texttt{iomr\_thr\_naive}}\\[-1pt]$0.0048$} & $\mathbf{0.41\times}$ \\
 & inflation & $13$ & $0.0019$ & \shortstack[r]{{\tiny\texttt{gmr\_ilp}}\\[-1pt]$0.0036$} & $1.90\times$ & \shortstack[r]{{\tiny\texttt{gmr\_bestofk}}\\[-1pt]$0.0155$} & $8.06\times$ \\
 & mixed & $13$ & $0.0080$ & \shortstack[r]{{\tiny\texttt{gmr\_ilp}}\\[-1pt]$0.0035$} & $\mathbf{0.43\times}$ & \shortstack[r]{{\tiny\texttt{iomr\_rand}}\\[-1pt]$0.0179$} & $2.24\times$ \\
\addlinespace[2pt]
\code{dimacs\_ny\_d} & deflation & $14$ & $0.0284$ & \shortstack[r]{{\tiny\texttt{l1sep\_gmr}}\\[-1pt]$0.0256$} & $\mathbf{0.90\times}$ & \shortstack[r]{{\tiny\texttt{gmr\_thr\_naive}}\\[-1pt]$0.0280$} & $\mathbf{0.98\times}$ \\
 & inflation & $14$ & $0.3191$ & \shortstack[r]{{\tiny\texttt{l1sep\_iomr}}\\[-1pt]$0.1649$} & $\mathbf{0.52\times}$ & \shortstack[r]{{\tiny\texttt{iomr\_thr\_naive}}\\[-1pt]$0.2049$} & $\mathbf{0.64\times}$ \\
 & mixed & $14$ & $0.1917$ & \shortstack[r]{{\tiny\texttt{spc\_iomr}}\\[-1pt]$0.1721$} & $\mathbf{0.90\times}$ & \shortstack[r]{{\tiny\texttt{iomr\_rand}}\\[-1pt]$0.2114$} & $1.10\times$ \\
\midrule
\multicolumn{8}{@{}l}{\textbf{topology: $k$-NN Jaccard ($\uparrow$; ${>}1$ is better)}} \\[2pt]
planted RGG & deflation & $14$ & $0.5973$ & \shortstack[r]{{\tiny\texttt{l1sep\_iomr}}\\[-1pt]$0.7721$} & $\mathbf{1.29\times}$ & \shortstack[r]{{\tiny\texttt{iomr\_bestofk}}\\[-1pt]$0.7284$} & $\mathbf{1.22\times}$ \\
 & inflation & $13$ & $0.8147$ & \shortstack[r]{{\tiny\texttt{gmr\_ilp}}\\[-1pt]$0.8003$} & $0.98\times$ & \shortstack[r]{{\tiny\texttt{gmr\_thr\_naive}}\\[-1pt]$0.5862$} & $0.72\times$ \\
 & mixed & $13$ & $0.6559$ & \shortstack[r]{{\tiny\texttt{gmr\_ilp}}\\[-1pt]$0.7845$} & $\mathbf{1.20\times}$ & \shortstack[r]{{\tiny\texttt{l1sep\_iomr}}\\[-1pt]$0.5250$} & $0.80\times$ \\
\addlinespace[2pt]
\code{dimacs\_ny\_d} & deflation & $14$ & $0.5896$ & \shortstack[r]{{\tiny\texttt{l1sep\_gmr}}\\[-1pt]$0.5915$} & $1.00\times$ & \shortstack[r]{{\tiny\texttt{gmr\_bestofk}}\\[-1pt]$0.5904$} & $1.00\times$ \\
 & inflation & $14$ & $0.4404$ & \shortstack[r]{{\tiny\texttt{spc\_gmr}}\\[-1pt]$0.4335$} & $0.98\times$ & \shortstack[r]{{\tiny\texttt{gmr\_bestofk}}\\[-1pt]$0.3984$} & $0.90\times$ \\
 & mixed & $14$ & $0.5044$ & \shortstack[r]{{\tiny\texttt{left\_edge}}\\[-1pt]$0.4931$} & $0.98\times$ & \shortstack[r]{{\tiny\texttt{iomr\_ilp}}\\[-1pt]$0.4494$} & $0.89\times$ \\
\bottomrule
\end{tabular}}
\end{table}

We run all repair algorithms on an RGG with $3{,}000$ vertices and on the metric \code{dimacs\_d\_ny} dataset, both corrupted with magnitude $3$, fraction $0.15$ on RGG and fractions $0.2,0.3$ for inflate, deflate (resp.) on \code{dimacs\_d\_ny}. The parameters differ so that the fraction of corrupted edges in each graph is roughly $0.15$. We then run the two downstream tasks and check the improvement compared to the observed graph. While the best performing algorithms improve almost all geometry tasks, they offer no relief for topology. Smaller solutions perform better, with the \GMR{} ILP obtaining the best results for $4$ of the instances. 

While the best performing algorithms paint an optimistic picture, they are measured against an \emph{oracle} of a known ground truth. In practice, one can expect an algorithm to perform like the median algorithm, reported on the right column. The picture is more bleak: Only one topology task showed improvement. A small set is good, but not enough.

\subsection{The repair values matter} \label{sec:weights_matter}
A random geometric graph $G = (V,E)$ comes equipped with \emph{true} Euclidean positions --- the positions of the sampled vertices. Let $P_{true}$ be those positions, and let $d^\star$ be the true Euclidean distances between the vertices in $V$. $G$ comes equipped with the weight function $d^\star$. Between non-adjacent vertices in $G$, the graph distance under $d^\star$ overestimates the true distance, so MDS run on $(G,d^\star)$ returns (different) positions $P_{d^\star}$. We write $\Phi_{d^\star}$ for the Procrustes disparity between $P_{true}$ and $P_{d^\star}$, this is a best case scenario; give the graph the true weights and try to recover the true positions. When we inject $G$ with a corruption, we observe the non-metric graph $(G,w)$. MDS run on $(G,w)$ returns positions $P_w$ that are further from the truth as compared to $P_{d^\star}$. Write $\Phi_{corrupt}$ for the disparity between $P_w$ and $P_{true}$. We then run some metric repair algorithm on $(G,w)$, and get a set $S$ of edges that we need to change to turn the graph metric. (Note that this set need not overlap the set of corrupted edges.) Instead of setting their weights according to \code{restore}, we set their weights to be \emph{the true weights}. We write $w^\star$ for this repair. Similarly, we write $\Phi_{w^\star}$ for the disparity between $P_{true}$ and $P_{w^\star}$. We now ask --- what fraction of the available repair does the set $S$ capture given the true distances:
\[
\mathrm{captured}(S) = \frac{\Phi_{corrupt} - \Phi_{w^\star}}{\Phi_{corrupt}- \Phi_{d^\star}}
.\]
The denominator is the gap in disparities between corruption and the original graph, while the numerator is the gap in disparities between a graph partially corrected with oracle access to true distances and the corrupted graph. The quantity is \emph{signed}, as $\Phi_{w^\star}$ can be larger than $\Phi_{corrupt}$; that is attempted repair can make the graph's embedding worse.
\begin{table*}[t]\centering\small
\caption{\textbf{Weights matter as well as the set.} A planted random geometric graph: $n = 300$, $m = 1{,}309$, $15\%$ of edges deflated $3\times$, planted corrupted set $|B| = 192$. Each row is a cover $S$; the two disparity columns reweigh $S$ by \restore{} versus the \emph{true} weights (lower is better).\emph{captured} is the share of the recoverable gap that the true weights close. }
\label{tab:cross}
\begin{tabular}{@{}lrrrrrrr@{}}
\toprule
 & & & & & \multicolumn{2}{c}{disparity after repair} & \\
\cmidrule(lr){6-7}
cover $S$ & $|S|$ & $|S|/m$ & precision & recall & \restore{} & true weights & captured \\
\midrule
\multicolumn{5}{@{}l}{\emph{observed }} & \multicolumn{2}{c}{$0.0388$} & $0\%$ \\
\multicolumn{5}{@{}l}{\emph{the clean graph}} & \multicolumn{2}{c}{$\mathbf{0.0033}$} & $\mathbf{100\%}$ \\
\midrule
oracle set $B$ & $192$ & $0.15$ & $1.000$ & $1.000$ & $0.0081$ & $\mathbf{0.0033}$ & $\mathbf{100\%}$ \\
\code{spc\_gmr} & $969$ & $0.74$ & $0.191$ & $0.964$ & $0.0783$ & $\mathbf{0.0049}$ & $\mathbf{95.4\%}$ \\
\code{pivot} & $705$ & $0.54$ & $0.258$ & $0.948$ & $0.1168$ & $\mathbf{0.0038}$ & $\mathbf{98.7\%}$ \\
\code{left\_edge} & $603$ & $0.46$ & $0.315$ & $0.990$ & $0.0384$ & $\mathbf{0.0058}$ & $\mathbf{92.8\%}$ \\
\code{domr} & $590$ & $0.45$ & $0.002$ & $0.005$ & $0.0388$ & $0.0388$ & $0.0\%$ \\
\code{spc\_iomr} & $390$ & $0.30$ & $0.485$ & $0.984$ & $0.0127$ & $\mathbf{0.0044}$ & $\mathbf{96.9\%}$ \\
\code{gmr\_rand} & $253$ & $0.19$ & $0.569$ & $0.750$ & $0.0127$ & $0.0147$ & $67.8\%$ \\
\code{gmr\_thr} & $253$ & $0.19$ & $0.569$ & $0.750$ & $0.0127$ & $0.0147$ & $67.8\%$ \\
\code{gmr\_bestofk} & $249$ & $0.19$ & $0.574$ & $0.745$ & $0.0127$ & $0.0147$ & $67.8\%$ \\
\code{iomr\_rand} & $243$ & $0.19$ & $0.642$ & $0.812$ & $0.0128$ & $0.0142$ & $69.3\%$ \\
\code{iomr\_thr} & $243$ & $0.19$ & $0.642$ & $0.812$ & $0.0128$ & $0.0142$ & $69.3\%$ \\
\code{iomr\_bestofk} & $237$ & $0.18$ & $0.658$ & $0.812$ & $0.0179$ & $0.0180$ & $58.6\%$ \\
\code{l1sep\_gmr} & $230$ & $0.18$ & $0.761$ & $0.911$ & $0.0112$ & $0.0114$ & $77.2\%$ \\
\code{l1sep\_iomr} & $230$ & $0.18$ & $0.761$ & $0.911$ & $0.0112$ & $0.0114$ & $77.2\%$ \\
\code{gmr\_ilp} & $181$ & $0.14$ & $0.867$ & $0.818$ & $0.0114$ & $0.0127$ & $73.5\%$ \\
\code{iomr\_ilp} & $181$ & $0.14$ & $0.928$ & $0.875$ & $0.0106$ & $0.0119$ & $75.7\%$ \\
\bottomrule
\end{tabular}
\end{table*}

For each repair algorithm and returned set $S$, we ask how much embedding disparity does it capture? \Cref{tab:cross} reports the results on an RGG with planted corruption. On sets that find the corruption ($|B|$, high recall), \restore{} over-corrects: \code{spc\_gmr} recovers $96\%$ of the deflated edges, yet \restore{} on its hitting set lands at $0.078$ --- worse than the observed $0.039$ --- while the true weights on the same edges close $95\%$ of the gap. The combinatorial heuristics \code{left\_edge} and \code{pivot} are simply large and rewrite the graph with the true weights, their recall is high while precision is low. The smallest sets obtained by the integer programs capture about $75\%$ of the recoverable gap, and do so in merely rewriting $0.14$ of the edges. Their recall is about $0.1$ away from the sloppy combinatorial methods, yet they improve precision by a factor of $3\times\sim 4\times$: The edges they miss cost $20\%$ of the fraction of available repair. At the same time, their disparity using \code{restore} is $3\times$ smaller than the combinatorial heuristics, so they do not lose as much information. Changing many edges pays off when you have oracle access to the true weights, being precise pays off if you use the naive rule. Both finding the correct edges and setting their weights impact the quality of recovery.

\section{Real data: performance and what repair recovers}\label{sec:realdata}\label{sec:down}

While synthetic data is sometimes assumed to represent the world, different datasets come from different scientific fields and consequently, their corresponding graphs differ in structure, density and non-metricity. We refer the reader again to Section~\ref{tab:datasets} for an overview of the family and their different properties. We emphasize that these datasets are used as \emph{probes} --- not all come from scientific fields in which the downstream tasks ($k$NN, MDS) we consider are run routinely. They are used to evaluate the \emph{algorithms} and the \emph{solution in the literature}, not the other way around. The one dataset on which MDS (or suitable variants) is routinely used is \code{nmr\_1d3z} and will be treated at the end of the section.
\subsection{The big picture}
We run all algorithms on all 19 real datasets,~\Cref{tab:sizegrid} holds the results. Much like \cref{sec:benchmarks}, no algorithm is a clear winner across the board. No dataset prefers \GMR{} or \IOMR{} algorithm for smaller hitting sets. Even on datasets where the best algorithms are for one problem variant, the difference between the best \GMR{} and best \IOMR{} is less than $1\%$, and no algorithm other than the ILPs (when they converge) is consistently good across all datasets. We note that \code{dimacs\_ny\_d} is a metric graph, so it is expected that all algorithms return a hitting set of size $0$.

\code{bct\_coactivation},\code{flycns\_male} and \code{fish1\_ten} are datasets in which the edge weights represent \emph{similarities}, rather than \emph{distance}: vertices that are similar should be considered as closer, not further apart, so we use order-inverting transformations to turn similarities into distances: the naive transformation takes a similarity $s$ to $1/s$. \code{\_lin} transforms the similarities linearly $s\mapsto s_{\max} - s + (10^{-9}s_{\max})$, \code{\_log} maps $s \mapsto \log(s_{\max}/s) + 10^{-9}$, and \code{\_raw} leaves weights unchanged so similarities are thought of as distances. The additive terms avoid $0$ weights. The choice of conversion impacts the non-metricity of the dataset and, as such, impacts the algorithms' performance: \code{\_lin} reduces non-metricity substantially and loses all non-metricity information in \code{fish1\_ten}. The extent of non-metricity is a \emph{design choice}, impacted by the transformation.

\subsection{Algorithms' performance in the wild}\label{subsec:perf_in_wild}
Five of our datasets are not metric by nature, but come equipped with \emph{external truths}: additional labels from the datasets that represent the true metric from which these graphs were generated, and their non metricity is an artifact of the measurements. \code{ripe\_atlas} and \code{dimacs\_ny\_t} come equipped with geographic distance between vertices, \code{pbmc3k} comes with the ambient cosine distance in the expression space, and \code{nmr} is equipped with the true protein fold and the Euclidean distance between vertices in the solved, true protein structure. 
\begin{table*}[t]
\centering
\caption{\textbf{On real data, repair recovers geometry once and topology never.} Per graph with an external truth, the best algorithm's $k$-NN neighbourhood recovery (topology, higher better) and MDS Procrustes disparity to the truth (geometry, lower better), against \code{observed}; \textbf{bold} clears a $0.5\%$ band. Geometry recovers on \code{dimacs\_ny\_t} alone ($-19\%$, editing $17$ of $6{,}017$ edges, via \code{spc\_iomr}); topology recovers nowhere, and \code{ripe\_atlas} ($95\%$ heavy) collapses. \DOMR{} is the control (Lemma~\ref{lem:noop}): its lift and disparity change are exactly $0$.}
\small
\setlength{\tabcolsep}{5pt}
\begin{tabular}{l r rll r rll}
\toprule
& & \multicolumn{3}{c}{topology ($k$-NN Jaccard, $\uparrow$)} & & \multicolumn{3}{c}{geometry (MDS disparity, $\downarrow$)} \\
\cmidrule(lr){3-5}\cmidrule(lr){7-9}
graph & $|H|/m$ & obs\,$\to$\,best & best algorithm & \% & & obs\,$\to$\,best & best algorithm & \% \\
\midrule
\code{ripe\_atlas} & 95.3\% & 0.428\,$\to$\,0.247 & \code{spc\_gmr} & -42.4\% & & 0.400\,$\to$\,0.416 & \code{spc\_gmr} & +4.0\% \\
\code{dimacs\_ny\_t} & 0.3\% & 0.562\,$\to$\,0.562 & \code{gmr\_bestofk} & -0.1\% & & 0.085\,$\to$\,0.069 & \code{spc\_iomr} & \textbf{-19.2\%} \\
\code{pbmc3k\_cosine\_knn} & 0.2\% & 0.696\,$\to$\,0.696 & \code{l1sep\_gmr} & -0.0\% & & 0.149\,$\to$\,0.148 & \code{gmr\_ilp} & -0.4\% \\
\code{nmr\_1d3z\_atom} & 1.1\% & 0.405\,$\to$\,0.404 & \code{gmr\_bestofk} & -0.4\% & & 0.149\,$\to$\,0.149 & \code{iomr\_bestofk} & -0.1\% \\
\code{nmr\_1d3z\_residue} & 5.2\% & 0.616\,$\to$\,0.597 & \code{gmr\_bestofk} & -3.0\% & & 0.105\,$\to$\,0.111 & \code{l1sep\_gmr} & +6.2\% \\
\bottomrule
\end{tabular}

\label{tab:realrecovery}
\end{table*}

We run all repair algorithms on these graphs and compare each repaired graph with the ground truth, both through topology and geometry.
Table~\ref{tab:realrecovery} reports the best answer for both probes on every graph with an external truth. The bottom line is that repair improves MDS once and $k$NN never.

\paragraph{What repair does to the $k$-NN graph}

Across every graph with an external truth, repair does not improve the $k$-nearest-neighbor structure: not a single algorithm beats the observed neighborhoods on any of the five, and typically deteriorates the structure. \DOMR{} does not change the corruption, since by~\ref{lem:domr} it does not
change the underlying distance function of $G$; every algorithm that actually reweighs an edge moves the neighborhoods the wrong way. How far wrong is governed by how non-metric the graph is: on the near-metric graphs (\code{dimacs\_ny\_t}, $|H|/m = 0.3\%$;
\code{pbmc3k}, $0.2\%$; \code{nmr\_1d3z\_atom}, $1.1\%$) the Jaccard index stays within half a percent,
\code{nmr\_1d3z\_residue} ($5.2\%$ heavy, $308$ edges) slips a few points, and \code{ripe\_atlas} collapses: with $95.3\%$ of its edges heavy, the least damaging reweighting drops the Jaccard index from $0.428$ to $0.247$ (\code{spc\_gmr}, $-0.18$) and most drive it to near zero ($-0.43$ for \code{spc\_iomr}).

On non-metricity that is inherent rather than planted, a reweighting pulls the distances toward
\emph{some} metric, not necessarily the true one, scrambling the local neighborhoods $k$-NN reads rather than restoring them. The graph-completers \code{pivot} and
\code{left\_edge} are the most destructive everywhere, and on the near-metric \code{nmr} graphs they are the \emph{only} algorithms that visibly hurt ($-0.084$ and $-0.075$ on the atom graph against $\approx -0.002$ for the targeted algorithms, $-0.18$ / $-0.20$ on the residue graph): completing to $\Theta(n^2)$ edges and reweighting changes the graph drastically, while a small hitting set --- the exact integer program, the covering-LP roundings --- leaves most of it in place. The ordering is the one the hitting-set size table already shows: the fewer edges an algorithm rewrites, the less of the $k$-NN structure it destroys. The full five-graph $\times$ three-$k$ grid is Table~\ref{tab:hope} (Appendix~\ref{app:extra}).

\paragraph{What repair does to MDS embedding}

In geometry, repair recovers exactly once. We embed each distance matrix with metric MDS, align it to the true configuration --- road geography, the folded structure, the expression space --- and report the Procrustes disparity, so lower is closer to the truth. On \code{dimacs\_ny\_t} every one of the fourteen algorithms beats the observed graph, and the best cuts the disparity from $0.085$ to $0.069$ (\code{spc\_iomr}, $-19\%$): a $0.3\%$-heavy graph over a genuine two-dimensional Euclidean truth, whose handful of real shortcuts are exactly
what pull the embedding off the map, so removing them pulls it back, and $1{,}724$ of its $6{,}017$ edges
are bridges that cannot violate the triangle inequality, shrinking the search further. Nowhere else does a
repair help: \code{ripe\_atlas} and \code{nmr\_1d3z\_residue} come out \emph{worse} at their best ($+4\%$,
$+6\%$), and \code{nmr\_1d3z\_atom} and \code{pbmc3k} remain the same. \DOMR{} is the fixed point on all five, the
same control as the topology axis.

\code{ripe\_atlas} is the opposite in every coordinate: $95.3\%$ of its edges are heavy, the non-metricity is a real property of latency rather than a geographic error, and a repair rewrites most of the graph to enforce a metric that is not geography. It does more than fail to help --- it makes the embedding \emph{less} Euclidean: the negative-eigenvalue mass climbs from $0.23$ to $0.36$--$0.42$ under repair, which Sonthalia, Van~Buskirk, Raichel, and Gilbert~\cite{gilbertMDS} prove bounds the classical-MDS error exactly, and which is indicative for the metric-MDS (\textsc{smacof})  disparity we report; repairing a graph toward metricity is not repairing it toward the plane. The \code{nmr} graphs sit between --- a three-dimensional truth from loose bounds that lowering a heavy edge to its detour cannot reconstruct, with the same completers \code{left\_edge} and \code{pivot} inflating the residue disparity to $0.33$ and $0.19$, two to three times the observed $0.105$. The ordering is again the hitting-set one: the smaller and more surgical the
set, the less of the true geometry it disturbs, and only where that geometry is a sparse, genuine, low-dimensional target does the repair move it the right way.


\subsection{NMR}
%
%
\begin{table*}[!tb]
\centering
\caption{\textbf{Weights and set matter, on real data.} The \restore{} column is the Procrustes disparity of the \restore{}-repaired fold to the deposited structure; the parenthesis gives it as a multiple of the observed disparity ($>1$ is worse). \emph{captured} (a percent) is the share of the available gain a cover buys when its edges are set to their \emph{true} distances. Rows run by cover size. Under \restore{} the best is \code{iomr\_bestofk} on \code{nmr\_atom} and \code{domr} on \code{nmr\_residue}, and on \code{nmr\_residue} \emph{none} of the 16 algorithms beats doing nothing. Only the 4 covers that rewrite 44--79\% of the graph capture anything. A representative subset is shown; the full per-algorithm grid is Table~\ref{tab:oraclefull} in Appendix~\ref{app:extra}.}
\small
\begin{tabular}{l rrrr @{\qquad} rrrr}
\toprule
& \multicolumn{4}{c}{\code{nmr\_atom}} & \multicolumn{4}{c}{\code{nmr\_residue}} \\
\cmidrule(lr){2-5}\cmidrule(l){6-9}
cover $S$ & $|S|$ & $|S|/m$ & \restore{} disparity & captured & $|S|$ & $|S|/m$ & \restore{} disparity & captured \\
\midrule
\itshape observed (no repair) & 0 & --- & 0.1489 & 0\% & 0 & --- & 0.1047 & 0\% \\
\itshape every edge true & 1{,}357 & 1.000 & --- & 100\% & 308 & 1.000 & --- & 100\% \\
\midrule
\code{gmr\_ilp} & 14 & 0.010 & 0.1497\,($1.01\times$) & -11.4\% & 14 & 0.045 & 0.1225\,($1.17\times$) & +1.9\% \\
\code{iomr\_bestofk} & 14 & 0.010 & 0.1487\,($1.00\times$) & -1.9\% & 16 & 0.052 & 0.1248\,($1.19\times$) & +1.1\% \\
\code{domr} & 15 & 0.011 & 0.1489\,($1.00\times$) & -8.5\% & 16 & 0.052 & 0.1047\,($1.00\times$) & -0.1\% \\
\code{l1sep\_gmr} & 14 & 0.010 & 0.1542\,($1.04\times$) & +15.4\% & 18 & 0.058 & 0.1112\,($1.06\times$) & -9.7\% \\
\code{spc\_gmr} & 44 & 0.032 & 0.1541\,($1.04\times$) & +10.6\% & 46 & 0.149 & 0.1263\,($1.21\times$) & -8.8\% \\
\code{pivot} & 1029 & 0.758 & 0.2164\,($1.45\times$) & +94.2\% & 136 & 0.443 & 0.2631\,($2.51\times$) & +44.2\% \\
\code{left\_edge} & 1071 & 0.789 & 0.1916\,($1.29\times$) & +104.2\% & 159 & 0.516 & 0.3270\,($3.12\times$) & +65.4\% \\
\bottomrule
\end{tabular}

\label{tab:oracle}
\end{table*}

\begin{figure*}[ht]
\centering
\caption{\textbf{Metric repair either wrecks the fold or does not change  it.} Three-dimensional MDS
of the \code{nmr\_residue} distance matrix, Procrustes-aligned to the deposited ubiquitin structure; the grey
ghost is the true fold, color is residue index, and all five panels share one axis box. \code{observed} is
what MDS on the observed graph produces. \code{domr} --- the \emph{best} of $\nmrNAlgoResidue$
algorithms --- is indistinguishable from it, because it does not affect the distance matrix.
Giving \emph{every} edge its true weight recovers most of the gap. 
\code{left\_edge} destroys the C-terminal tail.
}
\includegraphics[width=\textwidth]{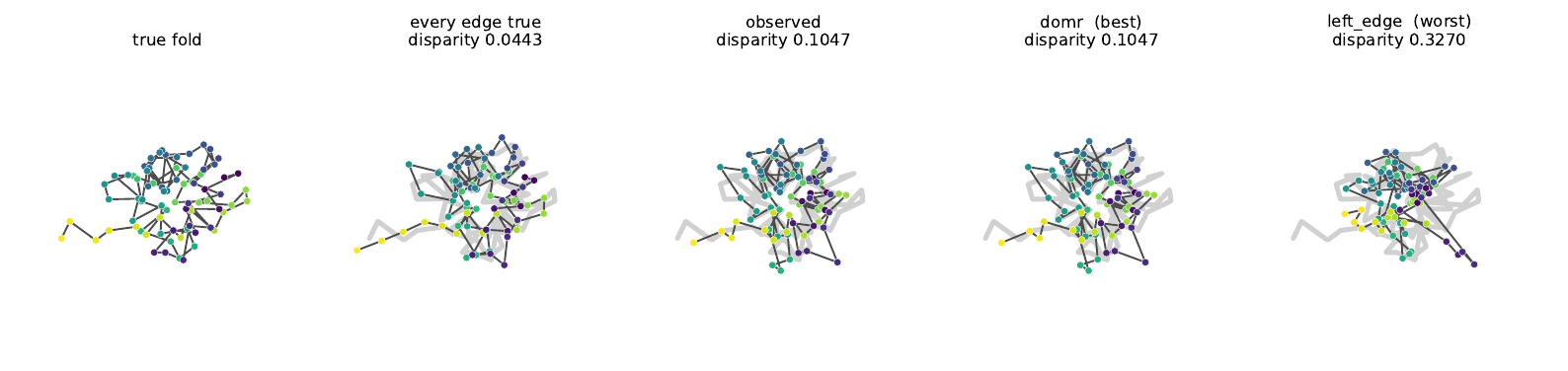}
\label{fig:nmrfold}
\end{figure*}
One graph is not a probe but the application itself, and we dedicate this part to it. Determining a protein structure by NMR \emph{is} a distance-geometry problem~\cite{nmr1,nmr2,nmr3,nmr5,nmr4}: the experiment reports \emph{upper bounds} of distance between vertices, and the task is to turn them into a three-dimensional fold. The embedding is the science, not a proxy we chose, and the non-metricity is inherent: the measured bounds, processed in various ways, routinely fail to satisfy the triangle inequality. MDS used to be the technique to determine the fold~\cite{crippenhavel1988,havel1998distance}.  While modern methods utilize techniques like simulated annealing and MCMC~\cite{nmr3}, MDS is still a part of the process. We are asking what a repair \emph{adds} to what every NMR pipeline already produces. 

NMR is also the only graph on which handing the true weights to the detected hitting set is well posed, as the true weights and measured weights are distances of the same unit.  Finally, the two graphs form a controlled pair. Both graphs represent the same protein and differ in one step: \code{nmr\_atom} keeps atom groups as nodes; \code{nmr\_residue} collapses each residue to a node under the \emph{minimum} bound over its atom pairs. The coarsening manufactures the non-metricity: a minimum invents a shortcut no single proton pair supports and raises $|H|/m$ by a factor of five. Embeddings are computed on the finite core ($\nmrCoreAtom$ of $\nmrNAtom$ atom nodes, all $\nmrCoreResidue$ residues).

\subsection{The best algorithm is the one that does nothing}

We embed each repaired distance matrix in three dimensions by metric MDS --- the classical embedding step of
distance geometry~\cite{liberti2014edg,crippenhavel1988} --- and measure the Procrustes disparity to the
deposited fold. This is not a probe, it is what MDS is \emph{for}.
Figure~\ref{fig:nmrfold} shows the result on \code{nmr\_residue}, with the true fold drawn as a gray ghost behind every panel so that a departure from it can be read directly.

 We rank the sixteen algorithms by the disparity of the repaired fold under \restore{} --- and the winner on
\code{nmr\_residue} is $\nmrBestCodeResidue$, at $\nmrRelDomrResidue\%$. It wins by not moving. \emph{Of the
$\nmrNAlgoResidue$ methods, $\nmrNWinResidue$ improve on doing nothing.} On \code{nmr\_atom} the situation is
no better in kind, only in degree: $\nmrBestCodeAtom$ leads at $\nmrRelIomrBestofkAtom\%$, a change of one part
in a thousand, and $\nmrNWinAtom$ of $\nmrNAlgoAtom$ beat the baseline --- all of them by less than a tenth of
a percent.

The cheap combinatorial methods are not close. $\nmrWorstCodeAtom$ rewrites $\nmrSMPivotAtom\%$ of the atom
graph and degrades the fold by $\nmrRelPivotAtom\%$; $\nmrWorstCodeResidue$ rewrites
$\nmrSMLeftEdgeResidue\%$ of the residue graph and makes the disparity \emph{triple},
$\nmrDispObsResidue \to \nmrRestLeftEdgeResidue$. Figure~\ref{fig:nmrfold} shows what that means physically:
the flexible C-terminal tail, which the truth extends and which \code{observed} still holds, is pulled into
the core and lost.

\subsection{The error is worth fixing}
\label{subsec:worth_fixing}
None of this would matter if the observed matrix were already as good as it gets, but it is not. As in~\Cref{sec:weights_matter}, we have access to a ground truth. \code{nmr} provides us with a deposited \emph{true protein fold positions} $P_{true}$ and the true distances $d^\star$. We think of \code{nmr\_residue} and \code{nmr\_atom} as observed corrupted graphs with weights $w$. We provide them with the true weights $d^\star$ and ask how close can the graph get to the truth. Set \emph{every} edge to its true distance in the residue graph and disparity compared to the true fold drops $\nmrDispObsResidue \to \nmrDispAllResidue$. In the atom graph, disparity drops $\nmrDispObsAtom \to \nmrDispAllAtom$. We once again measure the 
$\mathrm{captured}$ fraction by $S$

Table~\ref{tab:oracle} reports the captured fraction against $|S|/m$. The largest hitting sets \code{pivot} and \code{left\_edge} capture the most of the available repair, yet their \code{restore} disparity is the worst among all algorithms. They are not clever, they are simply large and rewrite a significant portion of the graph --- so true weights on their hitting sets will bring them closer to the truth. \code{left\_edge} captures \emph{more} than setting every edge its true weight on the \code{nmr\_atom} graph by hitting \emph{less} edges: we get more by changing less.  \code{l1sep\_gmr} tells the parallel story. By changing $1\%$ of the edges in the graph to their true weights, one captures $15\%$ more of the available recovery. A small set can make a large impact. However, not all small sets capture the same amount. \code{gmr\_ilp} is a smallest hitting set by design, and is of the same size as \code{l1sep\_gmr} but captures less than the baseline of no repair. At the same time, its \code{restore} disparity is lower than the one of \code{l1sep\_gmr}, so the choice of edges, their number, \emph{and} their assigned weights impact performance.

\section{Conclusion}\label{sec:conc}
Metric Repair has been studied as a combinatorial minimization problem: find the smallest set of edges that makes the graph metric, and an implicit assumption was that some corruption, either intentional or noisy, is responsible. However, this is hardly the whole story. 
Finding the smallest hitting set is the right \emph{first} question, further questions must specialize. The smallest set need not be the corrupted set even when increasing weights, and the type of corruption affects the algorithms' performance --- both direction and fraction. This is consistent across synthetic data and real data.

A repair is a set \emph{and} a weight assignment, and the rule \restore{} has been used in different fields for years. It is one option of many and it is rarely correct. Where an external truth exists, the hope that repair yields a better surrogate mostly fails: no topology is recovered anywhere in our study, and geometry only where the injury was large --- which $|H|/m$ predicts before anything is run. 

We study a specific type of corruption --- increase or decrease a fraction of the edges by a set magnitude. There are more corruptions that can, and should be studied, and might represent more families of real data: mixed-direction corruption or including edges that do not appear in the original graph to name a couple. We also point out that a theoretical analysis of why certain algorithms perform better under certain corruptions would be beneficial, alongside experimental work.

We close with what we consider the biggest question this study leaves. Finding a small set is not enough. One needs to find a combination of a correct small set, alongside a viable weight rule. A weight rule should be \emph{certified}, make the graph
metric; \emph{recovering}, move toward the truth; and \emph{oblivious}, use only the observed graph.
\restore{} is certified and oblivious; the true weights are certified and recovering; we know of nothing with
all three. Setting the weights is not an implementation detail. It is half the problem.

\clearpage
\bibliographystyle{plain}
\bibliography{refs}
\clearpage
\appendix
\section{Supplementary tables and figures}\label{app:extra}
\begin{table*}[t]
\centering\small
\caption{\textbf{The algorithm suite.} Every method benchmarked in the study, grouped by family; the \emph{variant} column marks whether it solves general (\GMR{}) or increase-only (\IOMR{}) metric repair. Three covering-LP rows are LP lower bounds that return no cover; only the integral \code{gmr\_lp\_rsp} yields a cover directly.}
\label{tab:suite}
\begin{tabular}{@{}llp{0.60\textwidth}@{}}
\toprule
algorithm & variant & description \\
\midrule
\multicolumn{3}{@{}l}{\emph{Exact solvers (integer program)}}\\[1pt]
\code{gmr\_ilp} & GMR & Exact minimum cover; lazy broken-cycle cut generation. \\
\code{iomr\_ilp} & IOMR & Exact minimum cover; lazy broken-cycle cut generation. \\
\midrule
\multicolumn{3}{@{}l}{\emph{Covering-LP relaxation and rounding}}\\[1pt]
\code{gmr\_lp\_rsp} & GMR & Covering LP with restricted-shortest-path (\code{rsp}) separation\\
\code{gmr\_lp\_naive} & GMR & Covering LP over canonical cycles only; LP lower bound (returns no cover). \\
\code{iomr\_lp\_naive} & IOMR & Covering LP; fractional LP lower bound (returns no cover). \\
\code{iomr\_lp\_rsp} & IOMR & Covering LP with \code{rsp} separation; fractional LP lower bound (returns no cover). \\
\code{gmr\_thr\_naive} & GMR & LP relaxation, deterministic threshold rounding. \\
\code{iomr\_thr\_naive} & IOMR & LP relaxation, deterministic threshold rounding. \\
\code{iomr\_thr\_rsp} & IOMR & LP relaxation (\code{rsp}), deterministic threshold rounding. \\
\code{gmr\_rand} & GMR & LP relaxation, randomized rounding. \\
\code{iomr\_rand} & IOMR & LP relaxation, randomized rounding. \\
\code{gmr\_bestofk} & GMR & LP relaxation, randomized rounding, best of $k$ independent draws. \\
\code{iomr\_bestofk} & IOMR & LP relaxation, randomized rounding, best of $k$ independent draws. \\
\code{iomr\_regiongrow} & IOMR & LP relaxation, region-growing (light-edge) construction. \\
\midrule
\multicolumn{3}{@{}l}{\emph{$\ell_1$ solver}}\\[1pt]
\code{l1sep\_gmr} & GMR & $\ell_1$ separation LP; iterative cut generation, then rounded to a cover. \\
\code{l1sep\_iomr} & IOMR & $\ell_1$ separation LP; iterative cut generation, then rounded to a cover. \\
\midrule
\multicolumn{3}{@{}l}{\emph{Combinatorial heuristics}}\\[1pt]
\code{spc\_gmr} & GMR & Greedy shortest-path cover~\cite{fan_et_al:LIPIcs.SWAT.2020.25}. \\
\code{spc\_iomr} & IOMR & Greedy shortest-path cover~\cite{fan_et_al:LIPIcs.SWAT.2020.25}. \\
\code{pivot} & GMR & Adapts \code{pivot} from~\cite{metricsultrametrics} to arbitrary graphs. \\
\code{left\_edge} & IOMR & Adapts the \IOMR{} algorithm from~\cite{gilbert2017sparse} to arbitrary graphs. \\
\midrule
\multicolumn{3}{@{}l}{\emph{Decrease-only}}\\[1pt]
\code{domr} & \DOMR{} & The \DOMR{} algorithm from~\cite{fan_et_al:LIPIcs.SWAT.2020.25}. \\
\bottomrule
\end{tabular}
\end{table*}

\subsection{The Algorithms}
We supply figures and tables that did not fit in the main text.
\subsection{Algorithms details}
We provide some implementation details. The algorithms \code{pivot} and \code{left\_edge} were designed for the complete graph $K_n$. A heuristic adaptation to arbitrary graphs $(G=(V,E),w)$ is to first add edges to $G$ until it is complete. If $uv\notin E$, we set its weight to $d_{G,w}(u,v)$. Then run the algorithm on the completion $K_n$ to obtain a hitting set $S$. The heuristic returns $S\cap E$. It is immediate to prove that $S$ is a hitting set on the completion $K_n$ implies $S\cap E$ is a hitting set on $G$.

For linear programs, we use a separation oracle in order to find unsatisfied constraints. The naive oracle simply finds a heavy edge $e$ and some shortest path between its endpoints to add as a constraint, this is the one used in \code{\_naive},\code{\_rand} and \code{\_best\_of\_k}. $\code{\_rsp}$ finds the \emph{optimal} constraint to add by solving the Restricted Shortest Path problem: If $x$ is a partial solution to the LP, the restricted shortest path problem asks a $u-v$ path $P$ that satisfies $w(uv) > w(P)$, and $\sum_{e\in P} x(e)$ is minimized. This can be solved in pseudopolynomial time~\cite{hassin1992approximation} and requires integer weights. We scale the weights of float-weighted graphs, and this solver typically times out. 
\begin{table*}[t]\centering\footnotesize
\caption{\textbf{Solution size on every real graph.} The fraction of edges each algorithm edits, $|S|/m$, on each real dataset; \textbf{bold} marks the smallest (best) and largest (worst) in each row, and \code{--} is no cover returned (timeout or non-convergence). $|H|/m$ is the heavy-edge fraction and $m$ the edge count. The covering-LP relaxations (\code{lp\_naive}) are lower bounds, not covers, and are omitted; \DOMR{} (the heavy set) leads the algorithm columns, so its entry equals $|H|/m$.}
\label{tab:sizegrid}
\resizebox{\textwidth}{!}{%
\begin{tabular}{@{}l rrr *{7}{r} | *{8}{r}@{}}
\toprule
& & & & \multicolumn{7}{c|}{\GMR{}} & \multicolumn{8}{c}{\IOMR{}} \\
\cmidrule(lr){5-11}\cmidrule(lr){12-19}
graph & $|H|/m$ & $m$ & domr & ilp & l1sep & spc & bestofk & rand & thr & pivot & ilp & l1sep & spc & bestofk & rand & thr & rgrow & left\_edge \\
\midrule
\code{nmr\_1d3z\_atom} & 0.011 & 1{,}357 & 0.011 & \textbf{0.010} & \textbf{0.010} & 0.032 & \textbf{0.010} & \textbf{0.010} & \textbf{0.010} & 0.758 & \textbf{0.010} & \textbf{0.010} & 0.021 & \textbf{0.010} & \textbf{0.010} & \textbf{0.010} & \textbf{0.010} & \textbf{0.789} \\
\code{nmr\_1d3z\_residue} & 0.052 & 308 & 0.052 & \textbf{0.045} & 0.058 & 0.149 & \textbf{0.045} & 0.052 & 0.052 & 0.443 & 0.052 & 0.062 & 0.114 & 0.052 & 0.052 & 0.052 & 0.052 & \textbf{0.516} \\
\code{dimacs\_ny\_d} & 0.000 & 6{,}017 & 0.000 & 0.000 & 0.000 & 0.000 & 0.000 & 0.000 & 0.000 & 0.000 & 0.000 & 0.000 & 0.000 & 0.000 & 0.000 & 0.000 & 0.000 & 0.000 \\
\code{dimacs\_ny\_t} & 0.003 & 6{,}017 & 0.003 & \textbf{0.002} & \textbf{0.002} & \textbf{0.011} & \textbf{0.002} & \textbf{0.002} & \textbf{0.002} & 0.003 & \textbf{0.002} & \textbf{0.002} & 0.008 & \textbf{0.002} & \textbf{0.002} & \textbf{0.002} & \textbf{0.002} & 0.003 \\
\code{ripe\_atlas} & 0.953 & 442{,}707 & 0.953 & \code{--} & \code{--} & \textbf{0.985} & \code{--} & 0.847 & 0.847 & \textbf{0.698} & \code{--} & \code{--} & 0.706 & \code{--} & 0.742 & \code{--} & \code{--} & 0.937 \\
\code{pbmc3k\_cosine\_knn} & 0.002 & 31{,}639 & 0.002 & \textbf{0.001} & 0.002 & 0.004 & \textbf{0.001} & \textbf{0.001} & \textbf{0.001} & 0.752 & \textbf{0.001} & 0.002 & 0.003 & \textbf{0.001} & 0.002 & 0.002 & 0.002 & \textbf{0.848} \\
\code{cassiopeia\_barcode\_knn} & 0.307 & 12{,}760 & 0.307 & \code{--} & \textbf{0.288} & 0.555 & \textbf{1.000} & 0.433 & \textbf{1.000} & 0.928 & \code{--} & \textbf{0.288} & 0.489 & \textbf{1.000} & 0.485 & \textbf{1.000} & \code{--} & 0.737 \\
\code{bct\_coactivation} & 0.435 & 18{,}625 & 0.435 & \textbf{0.113} & 0.161 & \textbf{0.666} & 0.169 & 0.169 & 0.169 & 0.352 & 0.119 & 0.138 & 0.256 & 0.155 & 0.155 & 0.155 & \code{--} & 0.352 \\
\code{bct\_coactivation\_lin} & 0.011 & 18{,}625 & 0.011 & \textbf{0.005} & \textbf{0.005} & 0.025 & \textbf{1.000} & \textbf{0.005} & \textbf{1.000} & 0.363 & \textbf{0.005} & \textbf{0.005} & 0.014 & \textbf{1.000} & \textbf{0.005} & \textbf{1.000} & \code{--} & 0.479 \\
\code{bct\_coactivation\_log} & 0.176 & 18{,}625 & 0.176 & \textbf{0.033} & 0.039 & 0.302 & \textbf{1.000} & 0.037 & \textbf{1.000} & 0.300 & \textbf{0.033} & 0.038 & 0.127 & \textbf{1.000} & 0.036 & \textbf{1.000} & \code{--} & 0.367 \\
\code{bct\_coactivation\_raw} & 0.181 & 18{,}625 & \textbf{0.181} & \code{--} & 0.885 & 0.365 & 0.436 & 0.443 & 0.443 & 0.585 & \code{--} & 0.885 & 0.877 & 0.841 & 0.843 & 0.843 & \code{--} & \textbf{0.905} \\
\code{flycns\_male} & 0.831 & 14{,}025 & 0.831 & \code{--} & \textbf{0.523} & \textbf{0.957} & 0.773 & 0.775 & 0.775 & 0.904 & \code{--} & \textbf{0.523} & 0.601 & 0.565 & 0.567 & 0.567 & \code{--} & 0.702 \\
\code{flycns\_male\_lin} & 0.006 & 14{,}025 & 0.006 & \textbf{0.000} & \textbf{0.000} & 0.012 & \textbf{1.000} & \textbf{0.000} & \textbf{1.000} & 0.013 & \textbf{0.000} & \textbf{0.000} & 0.006 & \textbf{1.000} & \textbf{0.000} & \textbf{1.000} & \textbf{0.000} & 0.211 \\
\code{flycns\_male\_log} & 0.126 & 14{,}025 & 0.126 & \textbf{0.016} & 0.020 & 0.186 & \textbf{1.000} & 0.020 & \textbf{1.000} & 0.265 & 0.017 & 0.019 & 0.064 & \textbf{1.000} & 0.020 & \textbf{1.000} & \code{--} & 0.430 \\
\code{flycns\_male\_raw} & 0.491 & 14{,}025 & \textbf{0.491} & \code{--} & 0.714 & 0.759 & 0.824 & 0.824 & 0.824 & 0.645 & \code{--} & \code{--} & 0.915 & 0.890 & 0.896 & 0.896 & \code{--} & \textbf{0.944} \\
\code{fish1\_ten} & 0.019 & 1{,}175 & 0.019 & \textbf{0.014} & \textbf{0.014} & 0.054 & \textbf{0.014} & \textbf{0.014} & \textbf{0.014} & 0.593 & \textbf{0.014} & \textbf{0.014} & 0.035 & \textbf{0.014} & \textbf{0.014} & \textbf{0.014} & \textbf{0.014} & \textbf{0.883} \\
\code{fish1\_ten\_lin} & 0.000 & 1{,}175 & 0.000 & 0.000 & 0.000 & 0.000 & 0.000 & 0.000 & 0.000 & 0.000 & 0.000 & 0.000 & 0.000 & 0.000 & 0.000 & 0.000 & 0.000 & 0.000 \\
\code{fish1\_ten\_log} & 0.003 & 1{,}175 & 0.003 & \textbf{0.001} & \textbf{0.001} & 0.008 & \textbf{1.000} & \textbf{0.001} & \textbf{1.000} & 0.846 & \textbf{0.001} & \textbf{0.001} & 0.004 & \textbf{1.000} & \textbf{0.001} & \textbf{1.000} & \textbf{0.001} & 0.881 \\
\code{fish1\_ten\_raw} & 0.020 & 1{,}175 & \textbf{0.020} & \code{--} & 0.021 & 0.061 & 0.030 & 0.032 & 0.032 & 0.031 & \code{--} & 0.042 & 0.100 & 0.051 & 0.054 & 0.054 & 0.054 & \textbf{0.454} \\
\bottomrule
\end{tabular}}
\end{table*}

We use two rounding schemes: Given an LP relaxation solution $x$, we either deterministically round up every entry that is above $1/f$, where $f$ is the length of the longest broken cycle in the graph, or round each $x(e)$ with probability proportional to its value. These are known rounding schemes.  \code{\_bestofk} draws $k$ independent roundings and picks the best one. 
\Cref{tab:sizegrid} reports the performance of all algorithms on all real datasets.
\subsection{Synthetic graphs experiments}
\begin{figure*}[t]\centering
\includegraphics[width=0.92\textwidth]{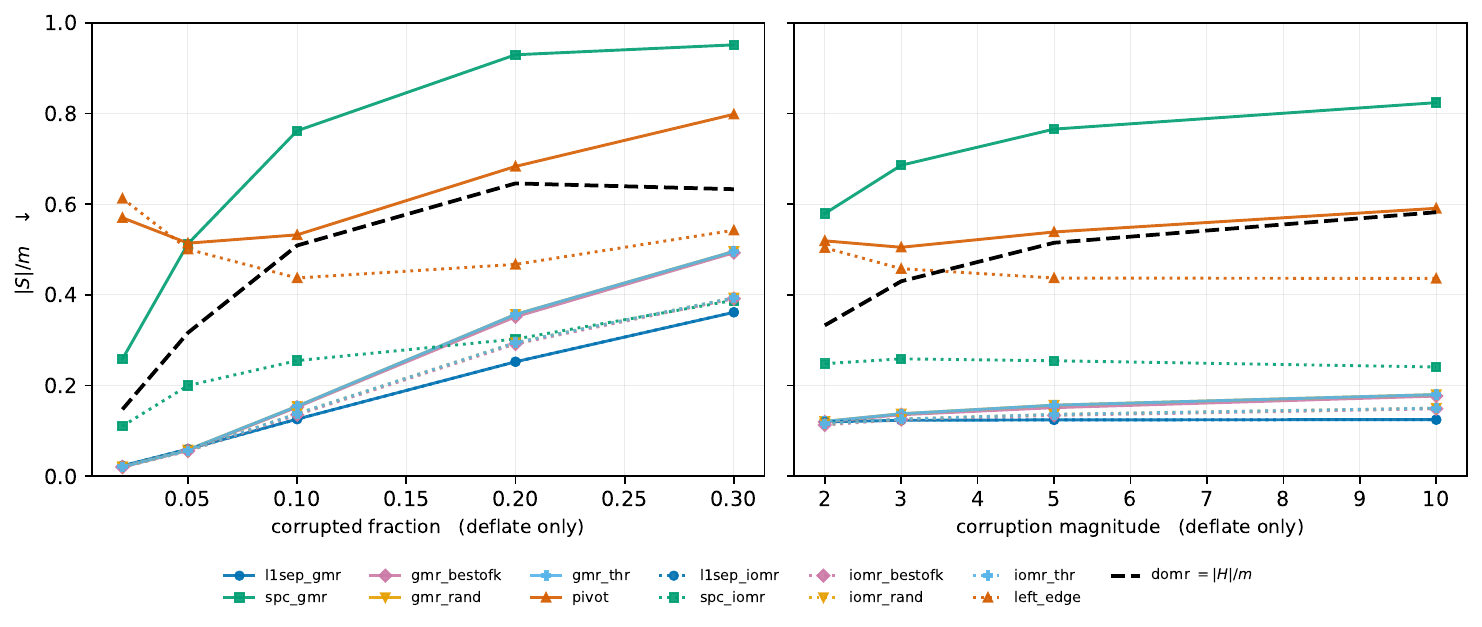}
\caption{\textbf{Fraction is a dial; magnitude is not.} $|S|/m$ against the corrupted fraction (left) and the
corruption magnitude (right). These sweeps carry only the \secFMDirs{} corruption; the direction inverts the
ranking, so they do not generalize to the other.}
\label{fig:fracmag}
\end{figure*}

\Cref{tab:optm} shows that measuring the algorithms' performance by $|S|/m$ does not move the ranking much compared to the true approximation ratio $|S|/\mathrm{OPT}$. Only \code{spc} algorithms shifted by more than $2$ positions, and it makes sense. They deterministically take more edges than they need.
\Cref{fig:fracmag} demonstrates that on RGGs with injected corruptions, the magnitude of the corruption is less indicative of solution size than the fraction of edges corrupted.
\begin{table*}[t]\centering\small
\caption{\textbf{$|S|/\mathrm{OPT}$ and $|S|/m$ rank the algorithms the same way wherever the graph has something to repair.} Median cover quality on the exact-ILP-converged instances, ordered by $|S|/\mathrm{OPT}$; the rank under each metric in parentheses, and the Spearman $\rho$ between the two orders per block. The two agree on the sparse RGG (both directions) --- which is why Section~\ref{sec:s52} reports $|S|/m$, defined at every scale, once the integer program stops converging. \code{pivot} ($\ast$) is the one heuristic where the two disagree by more than a place.}
\label{tab:optm}
\begin{tabular}{@{}lrr@{\quad}lrr@{}}
\toprule
\multicolumn{3}{c}{\shortstack{\emph{RGG inflation}\\[1pt]$\rho=0.92$}} & \multicolumn{3}{c}{\shortstack{\emph{RGG deflation}\\[1pt]$\rho=0.98$}} \\
\cmidrule(lr){1-3}\cmidrule(lr){4-6}
algorithm & $|S|/\mathrm{OPT}$ & $|S|/m$ & algorithm & $|S|/\mathrm{OPT}$ & $|S|/m$ \\
\midrule
\code{gmr\_ilp} & $1.00$\,(1) & $0.100$\,(1) & \code{iomr\_regiongrow} & $1.00$\,(1) & $0.011$\,(1) \\
\code{domr} & $1.00$\,(1) & $0.100$\,(2) & \code{gmr\_ilp} & $1.00$\,(1) & $0.097$\,(2) \\
\code{iomr\_ilp} & $1.00$\,(1) & $0.196$\,(3) & \code{iomr\_ilp} & $1.00$\,(1) & $0.098$\,(3) \\
\code{l1sep\_iomr} & $1.46$\,(4) & $0.253$\,(4) & \code{iomr\_bestofk} & $1.23$\,(4) & $0.121$\,(5) \\
\code{iomr\_bestofk} & $1.58$\,(5) & $0.272$\,(6) & \code{l1sep\_gmr} & $1.24$\,(5) & $0.120$\,(4) \\
\code{iomr\_rand} & $1.64$\,(6) & $0.282$\,(7) & \code{l1sep\_iomr} & $1.25$\,(6) & $0.122$\,(6) \\
\code{iomr\_thr\_naive} & $1.64$\,(6) & $0.282$\,(7) & \code{iomr\_rand} & $1.27$\,(7) & $0.125$\,(8) \\
\code{iomr\_regiongrow} & $1.66$\,(8) & $0.294$\,(9) & \code{iomr\_thr\_naive} & $1.27$\,(7) & $0.125$\,(8) \\
\code{spc\_iomr} & $1.97$\,(9) & $0.403$\,(13) & \code{gmr\_bestofk} & $1.27$\,(9) & $0.123$\,(7) \\
\code{spc\_gmr} & $2.59$\,(10) & $0.256$\,(5) & \code{gmr\_rand} & $1.32$\,(10) & $0.128$\,(10) \\
\code{gmr\_bestofk} & $3.12$\,(11) & $0.311$\,(10) & \code{gmr\_thr\_naive} & $1.32$\,(10) & $0.128$\,(10) \\
\code{gmr\_rand} & $3.38$\,(12) & $0.335$\,(11) & \code{spc\_iomr} & $2.59$\,(12) & $0.250$\,(12) \\
\code{gmr\_thr\_naive} & $3.38$\,(12) & $0.335$\,(11) & \code{left\_edge} & $4.58$\,(13) & $0.467$\,(14) \\
\code{left\_edge} & $3.45$\,(14) & $0.729$\,(16) & \code{domr} & $4.78$\,(14) & $0.391$\,(13) \\
\code{l1sep\_gmr} & $4.99$\,(15) & $0.494$\,(14) & \code{pivot}\,$\ast$ & $5.31$\,(15) & $0.505$\,(15) \\
\code{pivot}\,$\ast$ & $7.06$\,(16) & $0.689$\,(15) & \code{spc\_gmr} & $7.47$\,(16) & $0.648$\,(16) \\
\bottomrule
\end{tabular}
\end{table*}

\subsection{Real datasets experiments}
\begin{table*}[t]\centering\footnotesize\setlength{\tabcolsep}{4pt}
\caption{\textbf{The hope, answered.} The five real graphs carrying an external ground truth, with no planted corruption --- the data as they actually arrive. Topology is $k$-NN Jaccard against the truth ($\uparrow$), reported as the \textbf{lift} of the \emph{best} algorithm over the observed graph; geometry is the Procrustes disparity of an MDS embedding ($\downarrow$), again the best. Best, not median: the claim is negative, so repair is handed an oracle no practitioner could have (up to $15$ algorithms per cell, chosen \emph{with} the truth) and still loses. \textbf{Repair improves topology nowhere: not on one graph, at one $k$, by one algorithm --- not even the best one.} The one positive lift in the study ($+9 \times 10^{-6}$, \code{pbmc3k} at $k=10$) is an artefact: that graph \emph{is} the $15$-nearest-neighbour graph of its own truth, so recovery at $k \le 15$ is $0.999$ \emph{by construction}. Read it at $k=20$. \textbf{Geometry improves on $1$ graph of $5$} --- \code{dimacs\_ny\_t}, by $+19.2\%$. A win requires a $>\!1\%$ relative improvement. \DOMR{} is the control, not a competitor. On \code{ripe\_atlas} only $5$ algorithms return within the cap; on \code{nmr\_atom} the truth covers $343$ of the $430$ nodes.}
\label{tab:hope}
\begin{tabular}{@{}llrrrrrr@{\quad}rrlr@{}}
\toprule
 & & \multicolumn{2}{c}{$k=5$} & \multicolumn{2}{c}{$k=10$} & \multicolumn{2}{c}{$k=20$} & \multicolumn{4}{c}{geometry: Procrustes disparity ($\downarrow$)} \\
\cmidrule(lr){3-4}\cmidrule(lr){5-6}\cmidrule(lr){7-8}\cmidrule(l){9-12}
graph & what it is & obs. & best lift & obs. & best lift & obs. & best lift & obs. & best & by & gain \\
\midrule
\code{dimacs\_ny\_t} & road net (time) & $0.575$ & $-0.0012$ & $0.562$ & $-0.0008$ & $0.574$ & $-0.0002$ & $0.0854$ & $0.0691$ & \code{spc\_iomr} & $\mathbf{+19.2\%}$ \\
\code{nmr\_atom} & protein (atom) & $0.361$ & $-0.0060$ & $0.405$ & $-0.0015$ & $0.436$ & $-0.0000$ & $0.1489$ & $0.1487$ & \code{iomr\_bok} & $+0.1\%$ \\
\code{nmr\_res} & protein (residue) & $0.608$ & $-0.0325$ & $0.616$ & $-0.0188$ & $0.628$ & $-0.0171$ & $0.1047$ & $0.1112$ & \code{l1sep\_gmr} & $-6.2\%$ \\
\code{pbmc3k} & single-cell & $1.000$ & $-0.0024$ & $0.999$ & $+0.0000$ & $0.696$ & $-0.0002$ & $0.1488$ & $0.1481$ & \code{gmr\_ilp} & $+0.4\%$ \\
\code{ripe\_atlas} & internet latency & $0.343$ & $-0.1574$ & $0.428$ & $-0.1813$ & $0.473$ & $-0.1344$ & $0.3995$ & $0.4156$ & \code{spc\_gmr} & $-4.0\%$ \\
\bottomrule
\end{tabular}
\end{table*}

We show two more figures to complete the picture over the real datasets experiments. \Cref{tab:hope} expands on the experiment described in \cref{subsec:perf_in_wild}. The results for $k$NN remain consistent over different values of $k$. Under no $k$ and no dataset did the best lift get above $0$. That is, no repair algorithm made the computed $k$NN closer to the true $k$NN compared to the observed graph.

\begin{table*}[t]
\centering
\small
\begin{tabular}{l rrrr @{\qquad} rrrr}
\toprule
& \multicolumn{4}{c}{\code{nmr\_atom}} & \multicolumn{4}{c}{\code{nmr\_residue}} \\
\cmidrule(lr){2-5}\cmidrule(l){6-9}
cover $S$ & $|S|$ & $|S|/m$ & \restore{} disparity & captured & $|S|$ & $|S|/m$ & \restore{} disparity & captured \\
\midrule
\itshape observed (no repair) & 0 & --- & 0.1489 & 0\% & 0 & --- & 0.1047 & 0\% \\
\itshape every edge true & 1{,}357 & 1.000 & --- & 100\% & 308 & 1.000 & --- & 100\% \\
\midrule
\code{gmr\_ilp} & 14 & 0.010 & 0.1497\,($1.01\times$) & -11.4\% & 14 & 0.045 & 0.1225\,($1.17\times$) & +1.9\% \\
\code{gmr\_bestofk} & 14 & 0.010 & 0.1497\,($1.01\times$) & -11.4\% & 14 & 0.045 & 0.1125\,($1.07\times$) & +0.5\% \\
\code{gmr\_rand} & 14 & 0.010 & 0.1497\,($1.01\times$) & -11.4\% & 16 & 0.052 & 0.1268\,($1.21\times$) & +0.6\% \\
\code{iomr\_regiongrow} & 14 & 0.010 & 0.1487\,($1.00\times$) & -1.9\% & 16 & 0.052 & 0.1248\,($1.19\times$) & +1.1\% \\
\code{iomr\_ilp} & 14 & 0.010 & 0.1487\,($1.00\times$) & -1.9\% & 16 & 0.052 & 0.1230\,($1.18\times$) & +1.7\% \\
\code{gmr\_thr\_naive} & 14 & 0.010 & 0.1497\,($1.01\times$) & -11.4\% & 16 & 0.052 & 0.1268\,($1.21\times$) & +0.6\% \\
\code{iomr\_bestofk} & 14 & 0.010 & 0.1487\,($1.00\times$) & -1.9\% & 16 & 0.052 & 0.1248\,($1.19\times$) & +1.1\% \\
\code{iomr\_rand} & 14 & 0.010 & 0.1487\,($1.00\times$) & -1.9\% & 16 & 0.052 & 0.1248\,($1.19\times$) & +1.1\% \\
\code{iomr\_thr\_naive} & 14 & 0.010 & 0.1487\,($1.00\times$) & -1.9\% & 16 & 0.052 & 0.1248\,($1.19\times$) & +1.1\% \\
\code{domr} & 15 & 0.011 & 0.1489\,($1.00\times$) & -8.5\% & 16 & 0.052 & 0.1047\,($1.00\times$) & -0.1\% \\
\code{l1sep\_gmr} & 14 & 0.010 & 0.1542\,($1.04\times$) & +15.4\% & 18 & 0.058 & 0.1112\,($1.06\times$) & -9.7\% \\
\code{l1sep\_iomr} & 14 & 0.010 & 0.1542\,($1.04\times$) & +15.4\% & 19 & 0.062 & 0.1141\,($1.09\times$) & -9.9\% \\
\code{spc\_iomr} & 29 & 0.021 & 0.1542\,($1.04\times$) & +13.3\% & 35 & 0.114 & 0.1242\,($1.19\times$) & -10.2\% \\
\code{spc\_gmr} & 44 & 0.032 & 0.1541\,($1.04\times$) & +10.6\% & 46 & 0.149 & 0.1263\,($1.21\times$) & -8.8\% \\
\code{pivot} & 1029 & 0.758 & 0.2164\,($1.45\times$) & +94.2\% & 136 & 0.443 & 0.2631\,($2.51\times$) & +44.2\% \\
\code{left\_edge} & 1071 & 0.789 & 0.1916\,($1.29\times$) & +104.2\% & 159 & 0.516 & 0.3270\,($3.12\times$) & +65.4\% \\
\bottomrule
\end{tabular}
\caption{\textbf{Hand every hitting set the true weights, and the small ones still recover nothing.} The \restore{} column is the Procrustes disparity of the \restore{}-repaired fold to the deposited structure; the parenthesis gives it as a multiple of the observed disparity ($>1$ is worse). \emph{captured} (a percent) is the share of the available gain a hitting set buys when its edges are set to their \emph{true} distances. Rows run by set size. Under \restore{} the best is \code{iomr\_bestofk} on \code{nmr\_atom} and \code{domr} on \code{nmr\_residue}, and on \code{nmr\_residue} \emph{none} of the 16 algorithms beats doing nothing. Only the four hitting sets that rewrite 44--79\% of the graph capture anything. This is the full grid; the body shows the subset in Table~\ref{tab:oracle}.}
\label{tab:oraclefull}
\end{table*}

\Cref{tab:oraclefull} completes the picture of \cref{subsec:worth_fixing} and \Cref{tab:oracle}. It shows the full algorithmic suite run over the \code{nmr} dataset, the share of graph rewritten under the algorithm. We see the phenomenon described in \cref{subsec:worth_fixing} on a larger scale. Many algorithms that return small hitting sets capture different shares of the recoverable metric, and a few small solutions manage to capture a share fraction of the recoverable metric. It is not just about the edges, it is which edges and what weights.

\end{document}